
\input epsf
%
\input harvmac

%
%
\ifx\epsfbox\UnDeFiNeD\message{(NO epsf.tex, FIGURES WILL BE IGNORED)}
\def\figin#1{\vskip2in}
\else\message{(FIGURES WILL BE INCLUDED)}\def\figin#1{#1}\fi
\def\ifig#1#2#3{\xdef#1{fig.~\the\figno}
\goodbreak\midinsert\figin{\centerline{#3}}%
\smallskip\centerline{\vbox{\baselineskip12pt
\advance\hsize by -1truein\noindent\footnotefont{\bf Fig.~\the\figno:}
#2}}
\bigskip\endinsert\global\advance\figno by1}

\def\ifigure#1#2#3#4{
\midinsert
\vbox to #4truein{\ifx\figflag\figI
\vfil\centerline{\epsfysize=#4truein\epsfbox{#3}}\fi}
\narrower\narrower\noindent{\footnotefont
{\bf #1:}  #2\par}
\endinsert
}

\Title{RU-94-91}{\vbox{Lectures on Black Holes and Information Loss }}
\bigskip
\centerline{\it
T. Banks  \footnote{*}
{\rm  Lectures given at the Spring School on Supersymmetry, Supergravity,
 and Superstrings, Trieste, March 1994.
Supported in part by the Department of Energy under grant No.
DE-FG05-90ER40559  .} }
\smallskip
\centerline{Department of Physics and Astronomy}
\centerline{Rutgers University}
\centerline{Piscataway, NJ 08855-0849}
\noindent
\bigskip
\baselineskip 18pt
\noindent
In these lectures, the author's point of view on the problem of Hawking
Evaporation of Black Holes is explained in some detail.  A possible resolution
of the information loss paradox is proposed, which is fully in accord with the
rules of quantum mechanics.  Black hole formation and evaporation leaves
over a remnant which looks pointlike to an external observer with low resolving
power, but actually contains a new infinite asymptotic region of space.
Information can be lost to this new region without violating the rules of
quantum mechanics.  However, the thermodynamic nature of black holes can
only be understood by studying the results of measurements that probe
extremely small (sub- Planck scale) distances and times near the horizon.
Susskind's description of these measurements in terms of string theory may
provide an understanding of the Bekenstein-Hawking (BH) entropy in terms of
the states of {\it stranded strings} that cross the horizon.  The extreme
nonlocality of string theory when viewed at short time scales allows one
to evade all causality arguments which pretend to prove that the information
encoded in the BH entropy can only be accessed by the external observer
in times much longer than the black hole evaporation time.  The present author
believes however that the information lost in black hole evaporation is
generically larger than the BH entropy, and that the remaining information
is causually separated from the external world in the expanding horn of
a black hole remnant or {\it cornucopion}.  The possible observational
signatures of such cornucopions are briefly discussed.

\Date{October 1994}

\newsec{Introduction - The Facts In the Case}

The subject that I am going to talk about in these lectures, the Hawking
Evaporation of Black Holes, has been with us
for on the order of nineteen years now.  Although the last few years have seen
an upsurge of interest and activity in the subject, it remains a frustrating
field which stubbornly refuses to yield a satisfying resolution of its
paradoxes.
The number of very good physicists who have expressed fairly definitive
opinions
about the resolution of the Hawking puzzle is smaller than the number of
definitive opinions they have expressed.  The frustration is compounded
by the fact that there is no hope for experimental resolution of the confusion.
I would guess that there is still a sizable body of physicists and astronomers
who remain unconvinced of the observational evidence for the existence of
black holes.  Given that they exist, the probability that we are ever
going to examine a black hole close up seems very small.  Even if we could
examine
one close up the probability they we would happen to observe it at a time when
it was emitting substantial amounts of Hawking radiation is nil.  And most
frustrating of
all, even those physicists who insist that black hole radiation is not
 ``really''
incoherent, agree that it looks like thermal radiation for all practical
purposes.

Much of the recent activity and excitement has centered on two dimensional
models
in which it was hoped that one could examine the black hole puzzle in a
mathematically
controlled context.  Although this hope has so far proved illusory, I believe
that some progress has been made.  One is however cautioned that at least some
of the contributors to the field feel that these models do not capture
enough of the ``real'' problem to resolve the paradoxes.  This opinion is based
on the belief that the apparent paradoxes of Hawking radiation are pointing us
toward fundamental clues about the nature of quantum gravity.  Thus one may
argue
that only the full theory of the real world will give a satisfactory account
of the resolution of these paradoxes; models will not suffice.
I do not agree with this point of view, although as I will indicate below,
I am not completely in disagreement either.

The title of this section implies a certain degree of objectivity.  In a field
such as this, true objectivity is impossible.  I will therefore be presenting
``the facts'' in a way which emphasizes that part of the data that supports
my present opinions.  It is best then to get those opinions out on the table,
so
that you can judge for yourself what they are worth and how
much they are distorting my presentation of the facts.  In brief: I believe
that the
Hawking evaporation of black holes terminates in stable remnants.
An angular slice of the geometry
of those remnants is shown in Fig. 1.

\ifig\fone{An Angular Slice of the Static Geometry of a Black Hole Remnant}
{\epsfysize=6cm \epsfbox{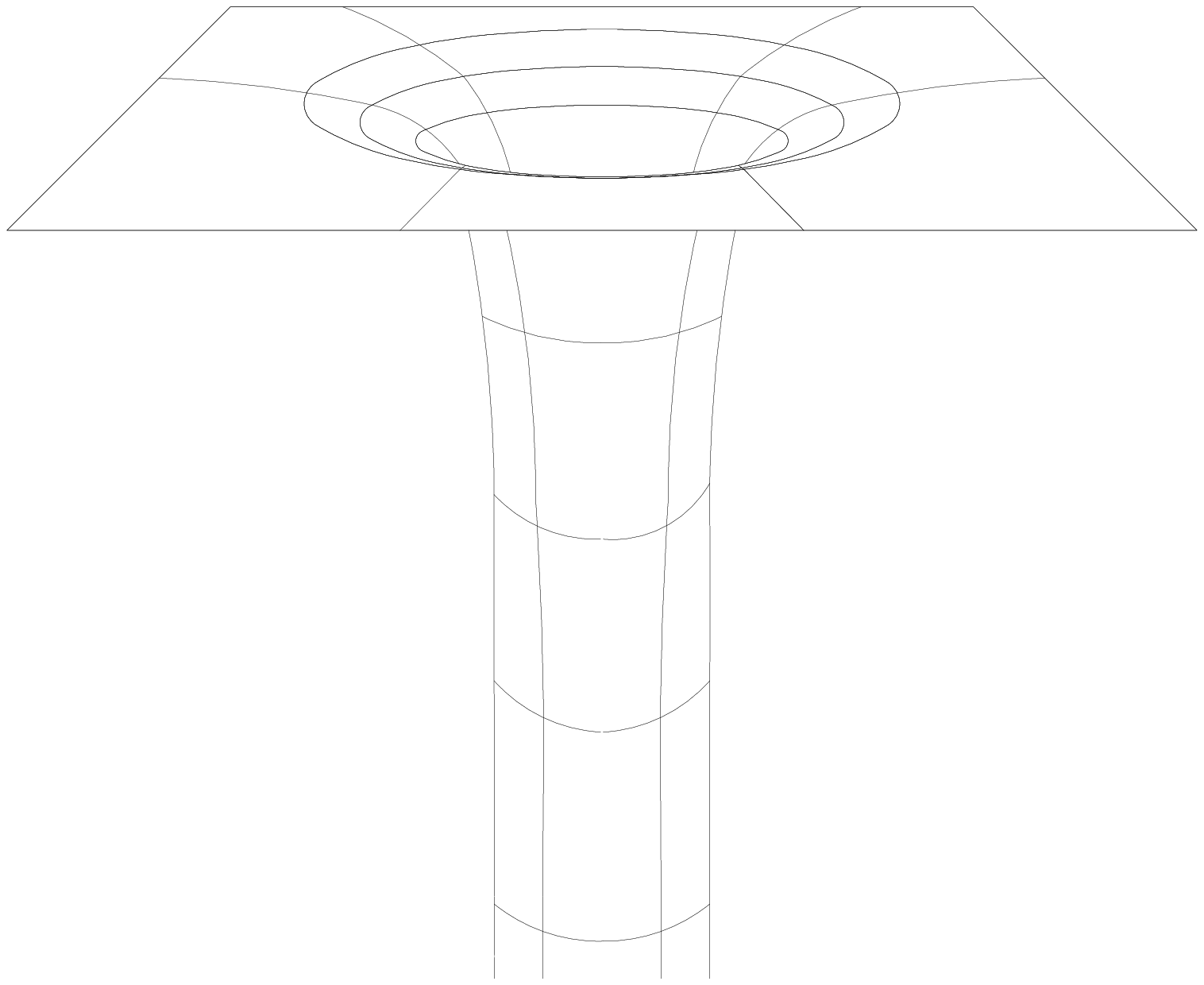}}

 It is a small ``hole in space'' attached
to a semi-infinite horn which has a spherical cross section of small radius.
These static geometries have a unique quantum ground state, but they are the
remains
of evolving geometries possessing a horizon which moved off to infinite
spacelike
distance.  The Penrose diagram of the full spacetime of one of these remnants
is
shown in Fig. 2.

\ifig\ftwo{The Penrose Diagram of a Black Hole Remnant}
{\epsfysize=6cm \epsfbox{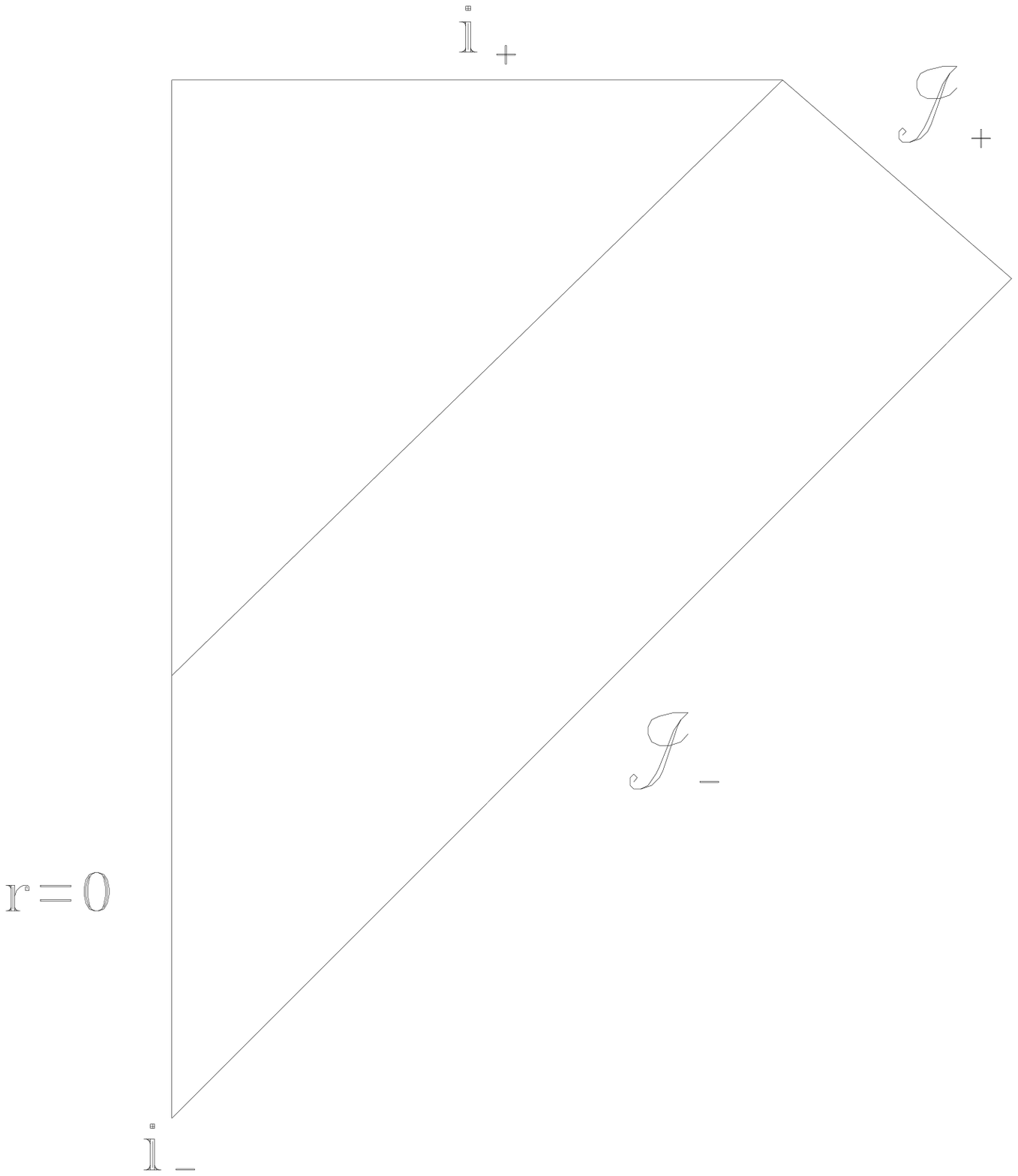}}

 There can be many different quantum states propagating behind the horizon
of these remnants, but the external geometry and ADM mass of all of these
states is
the same.  This is the repository for all of the information ``lost'' in
Hawking evaporation.
The full spacetime has (in the approximation in which quantum fluctuations of
the
gravitational field can be treated perturbatively) a unitary quantum mechanical
evolution, but the past and future contain different numbers of causally
separated
spatial asymptotic regions.  The S matrix for transitions from the initial
asymptotic
region back to itself is not unitary.

This by itself may seem like a resolution
of the information loss puzzle, and for a while I believed that it was.
However,
neither it nor any other discussion of information that is truly lost
to the external observer, can account
for the Hawking Bekenstein entropy.  This is a true thermodynamic entropy
which
describes interactions between the black hole and the external observer.
As such, it must be associated with degrees of freedom which are causally
connected to, and can interact with, the external observer.  By assigning the
entropy to correlations with degrees of freedom which are out of causal
contact with the external observer, or in a topologically disconnected
universe,
one gives up the possibility of explaining its thermodynamic nature.
I will argue in more detail below that the location of the
Bekenstein-Hawking degrees of freedom must be in an extremely tiny region in
the
vicinity of the horizon of the black hole.  In discussing them one is
inevitably
led into extremely short distance physics.  I believe that Susskind's string
model of these degrees of freedom, while still in a primitive stage of
development,
may lead to an ultimate explanation of the Bekenstein-Hawking entropy. If this
is the case, then the information represented by the BH entropy is not lost
to the external observer.

My picture of Hawking evaporation then, includes in some manner all of
the current theoretical prejudices about the subject.  I believe in
remnants of black hole evaporation, but all information stored in them
is causally inaccessible to the external observer.  The S-matrix for a
single asymptotic spatial region is not unitary once a black hole has
formed. However, the information whose absence is quantified by the
Bekenstein-Hawking entropy is not irretrievably lost.  It is located on
the horizon of the black hole, probably in the form of {\it stranded
strings} (see below for a definition), and will be radiated into the
original asymptotic region as the black hole evaporates.

\subsec{Some Classical Facts}

For the author at least, the only way to get intuition about what is going on
in general
relativity is to work in synchronous gauge.  This a name for any one of a
collection of coordinate systems in which ``time is time'', and general
relativity
is a theory of the dynamics of spatial geometry.  A synchronous gauge is chosen
by
picking a spacelike hypersurface and defining time to be the geodesic distance
orthogonal to this hypersurface.  For a typical hypersurface, synchronous gauge
may not be well defined on the entire spacetime manifold, but for the
Schwarzchild black hole there are two useful synchronous coordinate systems
that
tell us what is going on.  The simplest of these is related to internal Killing
coordinates by a simple reparametrization of the time variable.  The metric
takes the form
\eqn\killingmet{ds^2 = -dt^2 +   ({2GM\over\tau} -1)dr^2  +
\tau^2d\Omega^2}
\eqn\ttau{{t\over 2GM} = {\pi\over 2} - arccos(-\sqrt{\tau\over 2GM})
-\sqrt{{\tau\over 2GM}(1 - {\tau\over 2GM})}}
\eqn\ttautwo{{\tau\over 2GM}\sim ({t\over 2GM})^{2\over 3}\qquad :
t\rightarrow 0}
\eqn\ttauthree{0\geq {t\over 2GM}\geq -{\pi\over 2}; \qquad 0\leq \tau
\leq 2GM}
This coordinate system covers the inside of the Schwarzchild horizon.
Its equal time
surfaces are the hyperboloids with positive $UV$ in Kruskal coordinates.
  It is convenient for pictorial
purposes to modify the Schwarzchild metric by making the usual ``dust'' model
of
a collapsing star; sewing the geometry \killingmet onto a collapsing matter
dominated
Roberston Walker universe.  The time evolution of the geometry of the spatial
sections is shown in Fig. 3.

\ifig\fthree{The Spatial Geometry Inside a Schwarzchild Black Hole Evolves
Into an Infinite, and Infinitely Thin, Horn}
{\epsfysize=6cm \epsfbox{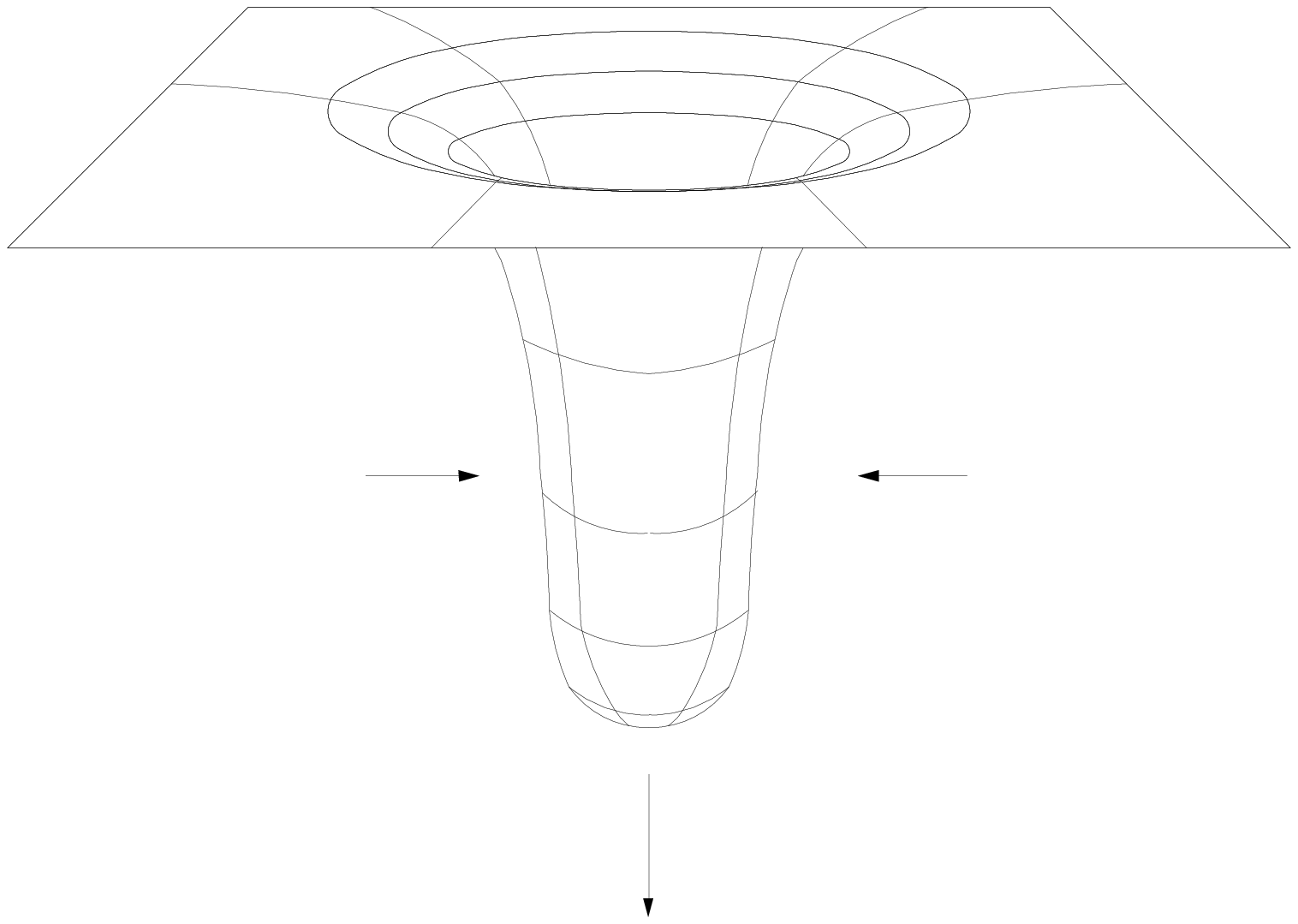}}

Note that the geometry has the ``hornlike'' shape
of Fig. 1, but the horn is dynamic.  It stretches along its length and shrinks
transversely, reaching infinite length and zero width in finite proper time.
This is the celebrated Schwarzchild singularity.  It has two aspects: the width
goes to zero, and the infinite length is achieved in finite time.  Both of
these
correspond to curvature singularities.

{}From these pictures, it is easy to understand why the black hole has a
horizon.
In Einstein's general relativity, space is allowed to stretch faster than
light can cross it.
A light beam sent out from some object may never be able to get back to its
point of origin
because in the time that it has travelled, the space has expanded (and
continues expanding).
Now let us introduce Novikov coordinates, a synchronous coordinate system
that takes
as its initial hypersurface the $t=0$ surface of external Schwarzchild
coordinates.
This coordinate system in fact covers the whole of the black hole (including,
though
we will not use this fact, its Kruskal extension).  Unfortunately the metric
has the
following ugly form in these coordinates
\eqn\novik{ds^2 = - dt^2 +{(\rho^2 + 1)\over rho^2} ({dR\over d\rho})^2
d\rho^2 + R^2 d\Omega^2}
where $R(\rho ,t)$ is a function defined by
\eqn\noviktwo{{t\over 2GM} = \pm(\rho^2 + 1)[{R\over 2GM} - {({R\over
2GM})^2 \over \rho^2 + 1}]^{\ha} + (\rho^2 + 1)^{3\over 2}
arccos[{R\over 2GM(\rho^2 + 1)}]^{\ha}}
Near $R=0$, which is the line ${t\over 2GM} = {\pi\over 2}
(\rho^2 + 1)^{3\over 2}$, ${dR\over d\rho}\sim R^{-\ha}$.
In Novikov coordinates the geometry of a collapsing star outside the horizon is
relatively slowly varying and becomes asymptotically static, while the inner
geometry
resembles that found in \killingmet .  It is clear then that if we look at
any point
a finite distance inside the horizon on some Novikov time slice, the space
between
it and any point outside the horizon will expand so rapidly that light will
not be able
to travel between them.  Indeed, in the Schwarzchild geometry, the radial
expansion rate
of the interior geometry in Novikov coordinates becomes infinite in finite
proper time.
Note however, that even if the singularity were absent, and the transverse size
of
the horn remained finite, we could still have superluminal expansion of the
interior
geometry and would be forced to conclude that the system had a horizon.

The synchronous view of the black hole interior suggests a possible scenario
for
its nonsingular evolution.  Namely, all that is necessary is to find a
mechanism
for stabilizing the horn against transverse collapse, and of slowing down the
expansion in the radial direction so that the rate never becomes infinite.
We will see below that this is precisely what happens for extremal charged
black holes.  One thing is clear about any such hypothetical mechanism for
stabilization of the black hole.  If we believe (as we should) that the
Schwarzchild solution
for a large mass black hole is valid down to times at which the transverse
dimension has shrunk
to microscopic size then the radial extent of the geometry will be very
large.
In fact, as we shall see, the example of charged extremal black holes
suggests that perhaps
the geometry keeps on growing in the radial direction, even when the
transverse collapse
is averted.  We are thus led to a picture of black hole final states as large
one dimensional protrusions on the geometry of space which connect onto the
space of the asymptotic
observer through essentially pointlike openings.  From many points of view,
the external
observer will regard these as point particles, but we will argue that this is
a mistake
when it comes to quantum mechanics.

We now turn to another well known classical feature of black holes, the
discrepancy
between the description of black hole physics given by infalling observers,
and those, like
the observer at infinity, who are supported in the gravitational field and
remain outside
the horizon.  When the black hole mass is large, the infalling observer
experiences
nothing in particular as she falls through the horizon.  To the asymptotic
observer
on the other hand, the horizon is a very peculiar place.  Nothing seems to fall
through it.  This can be ascribed to the behavior of the $g_{00}$ component
of the
metric in Schwarzchild coordinates .  Near the horizon the metric has the form
\eqn\rindler{- {x\over 2GM} dt^2 + {x\over 2GM}^{-1} dx^2 + {(2GM)}^2
d\Omega^2}
where $x = r - 2GM$.  The proper time per unit Schwarzchild coordinate time
goes to zero as
the horizon is approached.  This is a Lorentz contraction.  A supported
observer near the
horizon is accelerating like mad to prevent himself from falling through.
The instantaneous boost
relating his frame to that of the infalling observer is extremely large,
of order $e^{t\over 4M}$, where $t$ is the Schwarzchild time
coordinate.
For a large mass black hole, even moderate
energies will be boosted way above the Planck energy by such a transformation.
Thus the
Schwarzchild observer sees the infalling
observer as highly Lorentz contracted.  From his point of view the geodesic
observer's
clocks run very slowly and the structure in her machinery has very little
extent in the radial
direction .

Now suppose that the asymptotic observer has measuring apparatus of
limited accuracy, which cannot measure arbitrarily small intervals of space or
time.
Very quickly, the infalling observer reaches a point at which the full extent
of her
apparatus in the radial direction is apparently squeezed into less than a
minimally
measurable distance from the horizon.  All normal processes in the infalling
observers frame
are slowed down so much that the external observer cannot discern anything
changing with
time over the minimal interval between ticks of his clock.  Thus, a classical
observer
with limited powers of observation, quickly loses all information about what
is going
on in the infalling observer's frame.  The traditional ``covariant'' view of
relativity
is that this is a consequence of a bad choice of coordinates, but in the past
few years
a new paradigm has developed which takes the Schwarzchild observer's point of
view
as the basis of the treatment of the astrophysics of black holes, and their
interactions
with the external environment.  The Membrane Paradigm\ref\membrane{D.A.
MacDonald, R.H. Price, K.S. Thorne, {\bf The Membrane Paradigm}, Yale
University Press}, as it is called, claims that all
interactions of the black hole with the external world can be treated
correctly by a
model in which the black hole behaves {\it as if} there were a physical
membrane
with charge, current and energy densities on it hanging on a timelike surface
just above the horizon of the hole.  The classical aspects of this picture
can be {\it derived} from general relativity, but it has also been applied to
the
treatment of Hawking radiation.  We shall see that such a picture for
Hawking radiation seems to contradict the axioms of quantum field theory, but
that
it may be derivable from the dynamics of strings.

There are two important features of this classical picture that we will want
to remember later when
we discuss Susskind's conjectures about the nature of the Bekenstein-Hawking
Entropy.
The first is that the Schwarzchild observer's attempts to understand what is
happening to
his infalling colleague as she approaches the horizon require him to
contemplate measurements
of arbitrarily small length and time intervals.  Thus, a proper theory of
these measurements
requires us to understand physics at the shortest distances.  Secondly,
although
the Schwarzchild observer's picture is an ``incorrect'' picture of what is
``really happening''
to the infalling observer, it is a perfectly sensible account of everything
that the
Schwarzchild observer can actually measure.  In the end, I believe that this
will
be the sort of situation that we will recover for the quantum theory of black
holes, at least close to
the semiclassical limit of large mass.  Our picture of ``what is really going
on'' in
Hawking evaporation wil be the formation of a remnant and the disappearance of
particles behind a receding horizon.  However, the asymptotic observer will
be able to
account for much of what he observes in terms of a gas of ``stranded
strings'' glued to the
a membrane on the {\it stretched horizon}\ref\stretch{L.Susskind,
L.Thorlacius, R.Uglum, {\it Phys. Rev.}{\bf D48},(1993),3743.}.  I
will suggest below
that there is probably some real information loss to the asymptotic observer,
but that
the thermodynamic entropy of Bekenstein and Hawking represents information
that can
in principle (though certainly not in practice) be retrieved by him.

\subsec{Quantum Facts}

Here I will briefly review the salient facts about the theory of Hawking
radiation.
I assume that the listener/reader is already familiar with this material and
I will only
emphasize some important facts that are not usually presented.
Hawking's calculation
is carried out in the framework of an approximation to quantum gravity
called {\it quantum field
theory in curved spacetime}.  One imagines the formation of a black hole by a
classical
matter distribution falling in from infinity in an initially flat spacetime.
The first quantum correction
to this classical process consists of quantizing the linear fluctuations of
quantum fields around this
classical solution.  Hawking computed the S-matrix for this linear field
theory.  What does this involve?
The classical geometry has a perfectly well defined past asymptotic region.
The future however
consists of two causally disconnected asymptotic regions (the original one,
and a region ``down the horn''
in the synchronous gauge picture), one of which becomes singular a finite
time in the future.
Hawking's idea was to treat the singular region ``as if'' it had a well
defined set of asymptotic
states.  We can then consider {\it inclusive cross sections} in which the
external observer measures
only what is causally accessible to him, summing over the unknown final
states on the
other side of the horizon.  As usual this will lead to a density matrix
description of the
final state that he measures.

The expansion parameter for this semiclassical approximation is the Planck
mass divided by the mass of
the black hole.  In the limit of large mass the spacetime curvature is small
 (order $1\over (GM)^2 $)
everywhere except for the singularity\foot{In particular, it is small near
the horizon.}.
Furthermore, the singularity is a large timelike geodesic distance (of order
$GM$) from any point on the horizon.
In assessing the size of corrections to the semiclassical approximation one
must recognize that any quantum field
theory (and most particularly quantum gravity) has ultraviolet divergent
quantum fluctuations.
One must have in mind the value of the physical cutoff if one is to estimate
these.
There are two important possible choices for the cutoff scale.  One is the
Planck mass itself.  The other
possibility is to assume the existence of a small dimensionless parameter
which controls the
magnitude of the short distance fluctuations of gravity.  This is the
prescription of weakly coupled
string theory.  In effect the second option is the claim that the true cutoff
scale $L_S$, is larger
than the Planck length, the small dimensionless loop expansion
parameter being the square of the ratio $L_P \over L_S$.

According to Hawking's calculation\ref\hawk{S.Hawking,{\it Comm. Math.
Phys.}{\bf 43},(1975),199.}
 the expression for the outgoing density matrix is
\eqn\outdensity{\rho = e^{-{16\pi^2 GM H \over h}}}
where $h$ is Planck's constant.
The black hole is thus seen to behave like a black body with temperature
$T_H = {h\over 16\pi^2 GM} $.  Since its energy
is $M$, the first and second laws of thermodynamics give it an entropy
$S_{BH} ={8\pi^2 G M^2 \over h}  $
with the usual ambiguity of an additive constant.  Note that this entropy is
proportional to the
area of the horizon $ S_{BH} = {1\over 4}A M_P^2$.  According to the
Stefan Boltzmann Law, and taking into account that the area of the
horizon is $\sim {M^2}$ the black hole
should lose energy at a rate ${dM\over dt} = - {1\over M^2} $, giving it a
lifetime of order $M^3$.
All of this is for a neutral nonrotating black hole.  In general, the black
hole
temperature depends on its mass, angular momentum, and charge.
In particular, for near extremal charged black holes the lifetime is linear
in the deviation from extremality.

The paradox of all this is that the black hole seems to decay into incoherent
radiation.
Below we will review the argument that suggests that in standard quantum field
theory
the decay of the hole proceeds incoherently until a time when its energy
content is
very small compared to the amount of information that it has yet to liberate.
This appears
to lead to a choice between three alternative scenarios for the climax of the
radiation process,
all of which appear to lead to paradoxes.
We will enumerate and discuss them below.

The other key question raised by this discussion is the origin of the
Bekenstein-Hawking
entropy $S_{BH} ={1\over 4}A M_P^2 $.  It does not seem to come from a
counting of states.
In particular, if we try to compute the information theoretic entropy of
entanglement
in the density matrix computed by Hawking,
we find an infinite answer, coming from two different
sources.  One of these is silly.  We have computed the density matrix as if
the black hole
metric remained static for all time, even though it is emitting thermal
radiation.  This
gives us a trivial volume infinity.  The other infinity is more interesting,
and was first
discussed by 't Hooft\ref\entropy{G. 't Hooft, {\it Nucl. Phys.}{\bf B256},
(1985),727.}.  Its origin is in ultraviolet divergences.  Conventional
quantum field theory has an infinite number of states in any arbitrarily
small volume of space
time.  Furthermore, the correlation functions of degrees of freedom at nearby
points diverge
as the distance between the points goes to zero.  Thus if we choose a
spacelike surface and
make an imaginary line cutting it into two, the entropy of entanglement of
the degrees of
freedom on one side with those on the other, is
divergent\ref\srednicketal{M.Srednicki, {\it Phys. Rev. Lett.}
{\bf 71},(1993),666; B.DeWitt,{\it Phys. Rep.}{\bf 19C},(1975),297; \break
W.Unruh, N.Weiss, {\it Phys. Rev.}{\bf D29},(1984),1656;
L.Bombelli, R.Kaul, R. Sorkin, {\it Phys. Rev.}{\bf D34},(1986),373;
D.Kabat,M.Strassler,{\it Phys. Lett.}{\bf B329},(1994),46, hep-th 9401125;
C.Callan,F.Wilczek, {\it Phys. Lett.}{\bf B333},(1994),55,hep-th 9401072.}.
The coefficient
of this divergence depends on how we choose to cut off the theory.  't Hooft
and Susskind base
their theories of black hole evaporation on the contention that this infinite
correction
to the Hawking calculation of the entropy of the density matrix is a signal of
something wrong
in the conventional field theoretic treatment of the problem.  The goal of a
proper theory,
according to them, is to understand the physical origin of the BH entropy in
a context which allows one to compute the finite quantum corrections to the
BH formula.

\newsec{The Threefold Way}

\subsec{Subtle Correlations and Causality}

At first sight, the most conservative approach to the problem of information
loss
 is that which goes under the name of ``subtle correlations''.  According to
this dogma, the S-matrix for black hole formation and evaporation is unitary in
the Hilbert space of the original asymptotic observer.  The apparent loss of
coherence exhibited in Hawking's calculation is ascribed to the inadequacy of
his semiclassical approximation.  The standard analogy is to the heating
of a lump of coal:  Suppose that we encode the information in the Encyclopedia
Brittanica in Morse code and send it out as flashes of laser light that
are
 directed
 at a large lump of coal.  The laser flashes are absorbed by the coal,
which heats up.  All of the information in the Encyclopedia is now contained
in the coal.  Of course the heated coal emits infrared radiation and eventually
cools down.  After it has cooled, the information is stored in the heat that
it radiated, but for all practical purposes this radiation is thermal and the
information has been lost.

In this situation we understand what is going on.  The radiation from a
cooling
lump of coal is not really thermal, it is in a pure, albeit very complicated
, quantum state.  The useful information originally stored in the pulsed laser
beam is now encoded in correlations between photons which were emitted from the
coal at very different times, and are therefore very far from each other in
space.  This nonlocally stored information is of no practical use, and for
local
measurements, the pure state is equivalent to a mixed state.  Is this all
that
is going on in the Hawking calculation?

There is a very strong argument that this is not the case, at least not within
the
conventional formalism of quantum field theory.  The semiclassical
picture of Hawking evaporation is valid for most of the evaporation of a large
black hole\foot{This contention has been vigorously challenged by 't
Hooft\ref\hooft{G. 't Hooft,{\it Nucl. Phys.}{\bf B335},(1990),138.} and
others\ref\verlinde{E.Verlinde, H.Verlinde,{\it Nucl. Phys.}{\bf B406},
(1993),43.}.
The most compelling argument that it is correct comes from the large N
analysis of two dimensional evaporating black holes.  The large N
expansion
appears to provide a systematic renormalizable expansion of all
correlation
functions in the system which is controlled by a small parameter in the
entire
region of the semiclassical spacetime to the past of the spacelike slice
$99$
described below.  The expansion appears to break down only in a region
near
the singularity.}.  In particular, if we have an enormous black hole which has
evaporated away $99\%$ of its mass, leaving behind a hole which is still
large,
the Hawking calculation will be accurate to the past of the asymptotically
null spacelike slice labelled $99$ in the Penrose diagram of Fig. 4.

\ifig\ffour{Penrose Diagram Illustrating the Argument That Information Cannot
Get Out of a Black Hole Until Times Longer than The Hawking Lifetime}
{\epsfysize=6cm \epsfbox{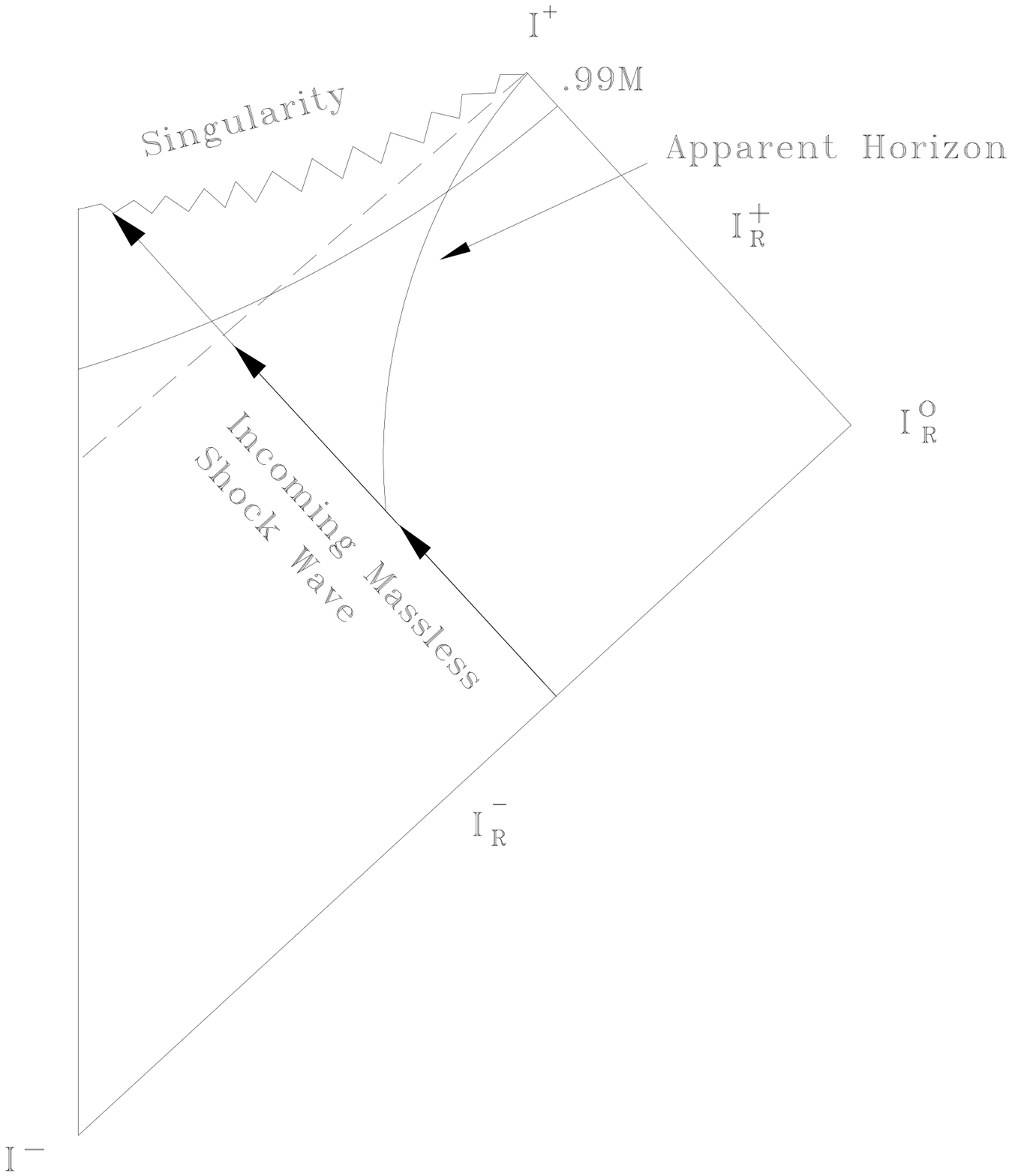}}

We can examine the state of that portion of the world which is behind
the
horizon.
In the semiclassical approximation, this will be calculable, and
correlated
with
the state of the outside world.  Observers falling through the horizon
, but still not at the singularity ,do not notice any particularly violent
events.  The interior state will be impure, with large entropy,
determined by the correlations (in the initial state) between objects
that fall into the black hole and those that don't. If we assume the
usual rules of quantum mechanics and locality, this will also be the
entropy of the state seen by an external observer.
There do not seem to be any principles which would prevent this entropy
of entanglement from being  of order the Bekenstein-Hawking entropy of
the initial hole, $\sim {M^2 \over M_P^2}$.
Thus, within the domain of validity of the semiclassical approximation, we can
establish the existence of a time slice on which the
energy of the external system is small $\sim .01 M$, but its entropy
huge, $\sim M^2$.

The argument above has been criticized on the grounds that for the asymptotic
observer, the time that passes between collapse and evaporation, is
large,
of order $M^3$.
  Small quantum corrections of order $1\over M$ could build
up over this period to (perhaps) ``fuzz out'' the horizon, and falsify our
conclusion that the information thrown into the black hole is causally
separated from
the external world on the time slice $99$.  It is for this reason
that we have concentrated on the state of the interior.  The proper time
interval that has elapsed since collapse is of order one on this part of
the
time slice.  Of course, the time slice $99$ that we have chosen is, in
this case, a rather strange one.  It hugs the light cone for much of its
extent.

Note that in two dimensions, and thus for near extremal charged black holes
for which much of the region of formation and evaporation is two dimensional,
the black hole's temperature is independent of its mass, and the evaporation
time is of order $M$.  This removes some of the above objections. An even
more rigorous justification of this argument can be found in the large
$N$
analysis of
two dimensional black holes that we will present in section 4.  There,
the
quantum
corrections are of order $1\over N$ while the time for evaporation of a
black hole
whose mass scales like $N$ is of order 1.  It seems to me that the only
way to avoid the conclusions of this argument is to give up the
assumption that the semiclassical approximation to quantum field theory
correctly describes physics in regions of spacetime where the curvature
is small.  Precisely such a retreat from the conventional wisdom is
proposed by the critics of this argument, though there seems to be no
agreement among them as to the correct replacement for semiclassical
quantum field theory.  I think it is fair to say that the above argument
establishes that the ``subtle correlation'' approach is far from
conservative.  It implies a radical rethinking of our approach to the
quantum dynamics of spacetime.

The only one of these radical
approaches that the author of the present lectures can make even a
pretense of understanding, is that of Susskind\ref\susstring{L. Susskind,
{\it Phys. Rev.}{\bf D50},(1994),2700, hep-th 9309145;
{\it ibid.}{D49},(1994),6606;{\it Phys.}\break
{Rev. Lett.}{\bf 71},(1993),2367.}.
Susskind's criticism of the above argument goes right to the
heart of quantum field theory.  Indeed, in making the above argument we
accepted without comment the assumption that the Hilbert space of the
whole system is a tensor product of states inside the horizon and states
outside.
This is the fundamental principle that allows us to conclude that information
cannot be radiated to the outside on the basis of the examination
of the states inside the horizon.
This assumption appears to be valid in quantum field theory and in naive
cutoff schemes
like lattice theories of gravity.  Susskind makes the point that it is in no
way valid in string theory.
If we cut space into two pieces, there will always be strings which straddle
the boundary.
It is to these ``stranded'' string states that Susskind looks for the origin
of black hole
entropy and the resolution of the paradoxes of Hawking evaporation.  We will
present a brief description of his work in section 6.

\subsec{The Remains of the Data}

Let us for the moment accept the argument given above, and examine its
consequences.
The fact that the external world has large entropy on the time slice
$99$
is already
in contradiction with the lump of coal analogy.  When a lump of coal has
radiated most of the energy it absorbed from the laser beam, and its glow
begins to fade, it has also radiated away most of the information in the
Encyclopedia.
We can go further however, and argue that the situation on the time slice $99$
implies the existence of a large set of degenerate long lived remnant states of
the evaporated black hole.  Suppose, in accord with the subtle correlation
hypothesis,
that the information trapped in the black hole on the time slice $99$
is later radiated into the external space\foot{This implies that our
theory
contains no
globally conserved quantities carried solely by massive particles.}
over a time $T$.  It is then contained in correlations between the late time
radiation and Hawking radiation which was emitted before $99$.  The
former is
contained
in a sphere of radius $T$ centered at the black hole\foot{For simplicity
we
are neglecting
the horizon radius of the black hole
of mass $.01 M$.} while the latter is mostly causally disconnected from this
 sphere.  The Hilbert space of the external world is , to a good approximation
a tensor product of the space of the late time radiation and that
emitted
before $99$\foot{Here again we invoke the tensor product structure of the
Hilbert space in local quantum field theory.}.
The state of the system is thus
\eqn\corrstate{\vert\Psi > = \sum \phi_{na}\vert a> \vert n>}
where $\vert a>$ is a state of the early time and $\vert n>$ a state of
the
late
time radiation.
The entropy of the state of the early time radiation:
\eqn\entropy{S = - \sum |\phi_{na} |^2 ln |\phi_{na} |^2 }
is supposed to be large, of order $M^2$.
Now consider the integral of the energy density over a sphere of radius larger
than $T$, but small enough to be spacelike relative to the region where most
of the early time Hawking radiation is.  This is, to good approximation,
a
positive
operator which acts only on the states $\vert n>$\foot{If there is a
hole in
this
argument, which I doubt, this is the most likely place for it to be
.  Radiation of large amounts of negative energy density might change our
conclusions.}.  Its expectation value is
\eqn\energy{<E> = \sum |\phi_{na} |^2 E_n < .01 M}
The only way to make \energy consistent with \entropy is to have many low
lying
energy levels $E_n$.  The late time radiation must therefore
occupy a large volume and the duration $T$ of the last stage of Hawking
evaporation
must be very long.
Thus, even if the information which fell behind the horizon of the black
hole
is
eventually returned to contact with the external observer, it must be stored
for a long time in a quasistable ``remnant''.   If we assume that the remnant
entropy is of order $M^2$, then the remnant lifetime implied by
the above argument is of order $M^4$, which for large black holes is
longer than the Hawking evaporation time, and typically longer than
the age of the universe.

The foregoing argument suggests that unless one envisions major
violations of
causality (or of the semiclassical treatment of gravity, which may be the same
thing)
in regions of spacetime where the semiclassical approximation shows no evidence
of breaking down\foot{Again I remind the reader that 't Hooft and others
have
argued that there is evidence for large corrections to the semiclassical
calculation, related to the fact that particles emitted as Hawking
radiation
have large Schwarzchild frequencies near the horizon.  I do not
understand
these arguments well enough to give a coherent account of them, but this
is
very likely a function of my stupidity rather than an intrinsic
deficiency of
the arguments.}, one must accept the existence of quasistable remnants.  Thus
even if black hole evaporation eventually returns all the information about the
collapsed system to the external observer, the actual process by which this
occurs is very different from the cooling of a lump of coal.  The ``unitary
S-matrix'', ``subtle correlation'' scenario reduces, for all practical
purposes, to the remnant scenario.

Further evidence for this contention comes from
the
investigation of Hawking radiation from moving mirrors by Carlitz and
Willey\ref\carlitz{R.Carlitz, R.Willey, {\it Phys. Rev.}{\bf D36},(1987),
2336; F.Wilczek, hep-th 9302096.}.
A moving mirror with a fixed trajectory (or indeed any fixed spacetime
geometry
 coupled to a quantum system but not satisfying self consistent
 equations with
 the expectation value of quantum stress as a source) cannot satisfy
 exact
energy conservation.  However, Carlitz and Willey find that if the mirror is
decelerated slowly enough to cause negligible violations of energy
conservation
, then the information that would have been lost if the deceleration had
not
occurred can only be returned over a very long time interval.  A similar
situation would obtain if we artificially reversed the collapse of the
interior of the Schwarzchild geometry, sometime after the spacelike
surface
$99$.  If we did this in a way that approximately preserves energy
conservation, we would find that the geometry could only return to flat
space
and the information return to the external observer, over a very long period.
This is particularly clear in the synchronous gauge description of black
hole geometry which we described in the introduction.
If we wait a long time before reversing the collapse, the spatial
geometry contains a large ``horn'' (see Fig. 1).  In order to shrink
this horn and return to a
flat spatial geometry without significant violation of energy
conservation, we require a long time.

I have emphasized Susskind's string theoretic
criticism of the basic tensor product structure of
the space of states in quantum field theory vitiates the force of these
arguments.
What I find hard to believe is that these criticisms significantly affect the
description of the evolution of a large black hole inside its horizon, long
before
the singularity is encountered.  Thus I believe that the picture of long
hornlike
geometries connected on to our space by tiny holes must be valid in string
theory as
well as in ordinary quantum field theory.
As we will see, consideration of such structures leads one to the notion
of stable remnants with infinite numbers of degenerate states, even if one
ignores
the above arguments.  Rather than being in contradiction with the existence of
remnants, I believe that Susskind's description of black hole
evaporation is like the membrane picture of classical black hole physics: if
it succeeds
it will provide a complete description of Hawking radiation and Bekenstein
Hawking entropy from
the point of view of the external observer.  It does not contradict the
alternate
description of physics as seen by an infalling observer.  In the final section
we will make some
comments about the differences between these two points of view.  Their
reconciliation relies on the distinction between information theoretic and
thermodynamic entropy in situations in which different parts of the universe
 become causally separated from each other.

We turn now to an exposition of the
apparent phenomenological catastrophe caused by the hypothesis of stable
or
quasi
stable black hole remnants.
Hawking's calculation of the evolution of an evaporating black hole appears to
be valid until the mass of the hole is of order the Planck mass.  Its
Schwarzchild radius is then of order the Planck length, and it appears
pointlike
to all but the most well equipped external observers.  The remnant scenario
thus appears to require the existence of a new class of ``particles''
all of which have masses of order the Planck mass.  On the other hand,
there must be a distinct remnant particle,
for each possible state of matter that can collapse to form a black
hole, and since remnants can be formed starting with black holes of
arbitrarily
large mass , there must be an essentially infinite number of different
remnant species.
Even a few species of stable Planck mass particles might cause
difficulties for cosmology if they are produced with reasonable probability
after inflation, but an infinite number of species is a complete disaster.
Schwinger's calculation of the pair production of charged particles in
a background electric field show that the probability depends only on
the mass
and charge of the particle (for fixed field strength).  An infinite
number of
degenerate charged species would give an infinite cross section for this
process.  (Similarly, the formula for Hawking production of a particle
around a
fixed black hole depends only on its mass.)  We apparently stand in
danger of
producing an infinite shower of black hole remnants every time we turn
on the
lights.  One of the most important ideas discussed in this review will be the
demonstration that these estimates of remnant production are completely wrong,
because black hole remnants do not behave like elementary particles even though
they look pointlike to an external observer.  This argument will be
taken up
in the next section.

\subsec{Information Loss}

First however we must review the third major scenario for the endpoint
of
Hawking evaporation, that proposed by Hawking himself.  Eschewing both
the
"information is returned in subtle correlations" approach and the idea
of
remnants (essentially for the reasons outlined above) Hawking instead
proposed that an evaporating black hole simply disappeared, taking with it
the information that was lost to the external observer in the collapse process.
For very small black holes the entire process of formation and
evaporation
occupies a small region of space time.  Since it is (in Hawking's view)
a
completely local phenomenon, it should happen all the time in the form
of
virtual processes even when sufficient energy for real black hole
formation is
unavailable.  Again due to the locality of the process, we should be
able to
construct a coarse grained, or effective, theory, describing the effect of
these virtual information destroying processes on large scale physics.
Hawking indeed proposed a formalism for computing such corrections.
Since pure states can now evolve into mixed states the effective theory
must
now map density matrices to density matrices in a way which does not
preserve
purity.

In ordinary quantum mechanics, the initial and final density matrices in
a
scattering experiment are related by
\eqn\unitarity{(\rho_{out} )_A^B  = {S_A}^C {(\rho_{in} )_C}^D
{S^{\dagger}_D}^B}
where $S$ is the scattering matrix.  Hawking proposed instead a general
linear
relation\foot{Nonlinear density matrix evolution equations lead to
nonlocal
phenomena which Polchinski\ref\polchinski{J.Polchinski, {\it Phys. Rev. Lett.}
{\bf 66},(1991),397.} has dubbed "Everett Phones".
EPR correlations can be used to send messages in such theories.
It is not clear to exactly what extent this is ruled out by experiment,
but we will not discuss nonlinear density matrix evolution in this review.}
\eqn\dollarmat{(\rho_{out} )_A^B = (\$ )_{AC}^{BD} (\rho_{in} )_D^C}
If the $\$$ matrix factorizes into the product of an $S$ matrix and its
inverse, then we have unitary evolution, preserving purity.  If it does
not
so factorize then purity is lost.  Hawking proposed\ref\dolarmat{S.Hawking,
{\it Phys. Rev.}{\bf D14},(1976),2460.} that
the
true $\$$ matrix of the world had a small, nonfactorizable term of the form
\eqn\nonfact{\delta (\$ )_{AB}^{CD} = C_{ij} ({1\over M_P^p}\int O_i\int O_j}
where the $O_i$ are operators of high dimension, as indicated by the
powers of
the Planck mass.

Unfortunately these apparently small corrections to the $\$$ matrix are
not small at all.  To see this\ref\bps{T. Banks, M.Peskin, L. Susskind,
{\it Nucl. Phys.}{\bf B244},(1984),135.} we will have to make a small
extension of
Hawking's $\$ $ matrix formalism and discuss local time evolution
equations
for the density matrix.  That this must be possible follows from the
assumption that processes of virtual formation and evaporation of small
black
holes are, in Hawking's picture, confined to a small space time region.
With sufficient coarse graining, we must be able to incorporate their
effect
in a set of local evolution equations.  We will see that even if we take
the
coarse graining scale to be a nuclear time scale, the "small" terms in
the
evolution equation analogous to \nonfact are far from negligible.
Let us begin by writing the most general linear coarse grained evolution
equation for the density matrix:

\eqn\evolut{\dot{\rho}_A^B = {\cal H}_{AC}^{BD}\rho_D^C}

We now impose the conditions that $\rho$ remain hermitian and that
probability
be conserved.  To do this it is convenient to expand the four index object
${\cal H}$ as a sum of tensor products of matrices.  We then find the
most
general probability and hermiticity conserving linear equation to
be\ref\lindblad{G.Lindblad, {\it Comm. Math. Phys.}{\bf 48},(1976),119.}
\eqn\evoluttwo{\dot{\rho} = i [H,\rho ] + C_{\alpha\beta}[O_{\alpha},
[O_{\beta}, \rho ]]}
where $H$ is hermitian and $C_{\alpha\beta}$ is a real matrix.
The $O_{\alpha}$ run over a complete set of hermitian operators.
To preserve the positivity of $\rho$ we must impose a condition on the
relative sizes of the symmetric and antisymmetric parts of $C$.
We {\it might} also want to impose a condition guaranteeing that entropy
always increases \bps.

When $C$ is a symmetric matrix, it is possible to make a simple model which
produces the equation \evoluttwo .  It is simply quantum mechanics
coupled to
random sources via a hamiltonian
\eqn\randsource{H_R = H + J_{\alpha}(t) O_{\alpha}}
where the $J$'s have white noise correlation functions
\eqn\corrfcn{< J_{\alpha}(t) J_{\beta}(s)> = C_{\alpha\beta}\delta (t - s)}
This interpretation makes most of the important features of
the equation \evoluttwo obvious.  In particular, although the evolution
equation is time translation invariant, it does not conserve energy.
Time
translation invariance guarantees only the conservation of the average energy.
In a random system there will generically be energy fluctuations and the
moments of the energy will not be preserved.  Similarly, space
translation
invariance of \randsource does not guarantee conservation of momentum.

The extent of this violation of the conservation laws depends on the
extent to which the operators $O_{\alpha}$ and the correlation function
of the sources are local.  We have assumed that we are working at a
scale for which the time correlation of the sources is local.  Hawking's
proposals lead one to expect all the nonlocality in the new terms in
the equation to be at the Planck scale.  As shown in \bps this leads to
disaster. The inverse powers of the Planck length are cancelled by
matrix elements of the local operators between states of very low and
very high energy.  In a flash, the vacuum is converted into a mixed
state whose dominant components have very high energy.  To make the
violations of purity small we have to smear the operators over long
distance scales, which leads to violations of locality.

Special examples exist, in which the non quantum mechanical terms in the
evolution equation involve only conserved charges.  However, all such
examples violate either the positivity of the density matrix, or Lorentz
invariance \ref\sredpurity{M.Srednicki, {\it Nucl. Phys.}{\bf B410},(1993),
143, hep-th 9206056.}. It is also unclear how such nonlocal equations
for the density matrix could arise from integrating out processes which
are localized in space and time.

It has been argued by Hawking\ref\hawkrebut{S.Hawking, {\it Nucl. Phys.}
{\bf B244},(1984),135.}, that these arguments must
be wrong.  Adopting Dyson's \ref\dyson{F.Dyson, Institute for Advanced Study
preprint, 1976, {\it Unpublished}.}idea that information loss is
due to the splitting off of a disconnected universe, Hawking argues that
such a process must conserve energy and momentum.  This argument is
correct, but overly classical.  In a generally covariant quantum theory,
the amplitude for splitting off
a disconnected universe must be summed coherently over all points of
spacetime.  Both the time and place where the splitting occurs, are
infinitely uncertain.  The result of integrating out such processes, is
an effective action which is infinitely nonlocal in time as well as
space.  In the theory of wormholes
 it is shown that the sum
over topologies can be rewritten as a theory in which
the constants of nature are uncertain, but once they are determined by
experiment,
the local dynamics of the universe is described by quantum mechanics
\ref\worm{S.Coleman, {\it Nucl. Phys.}{\bf B307},(1988),864; S.Giddings,
A.Strominger, {\it Nucl. Phys.}{\bf 307},\break (1988),354;
 T. Banks, {\it Nucl. Phys.}{\bf B309},(1988),643, {\it Physicalia}{\bf 12},
(1990),19.}.
Splitting off of disconnected universes
does not lead to evolution equations for the density matrix of the
kind studied in \bps , and cannot describe the spatially and temporally
localized process of black hole formation and evaporation.

I should not conclude this section without mentioning a possible loophole in
the argument.  There is nothing that prevents us from writing a nonquantum
mechanical evolution equation for the density matrix that has an extremely
small
dimensionless parameter in front of the double commutator term.  The
coefficient
is severely constrained by the long lifetimes of metastable nuclei that alpha
decay
via quantum tunneling processes, but no experiment can constrain a coefficient
to be exactly zero.  A theory of information loss which naturally explained
the small dimensionless parameter, would be consistent with experiment.

An interesting model which attempts to incorporate this idea has been
constructed recently by Strominger\ref\andymystery{A.Strominger, hep-th
 9405094.}.  He modifies the
RST\ref\rst{J.Russo, L. Thorlacius, L.Susskind, {\it Phys. Rev.}{\bf D47},
(1993),3444.} model of two
dimensional black hole evaporation by treating the model as an
open string theory with a string coupling that depends on the values of
world sheet fields in an unusual manner\foot{That is, he considers open
string theory with a singular spacetime dilaton (not to be confused with
the world sheet dilaton$\!$) field.}.  The string splitting and joining
interaction
vanishes unless the world sheet dilaton field takes on the critical
value corresponding to world sheet black hole formation.  In the
semiclassical, large $N$ limit, this only occurs near the worldsheet
position of black hole formation.  The probability of its occuring
anywhere else is suppressed by a factor $e^{-N}$. This model
then appears to be an example of true information loss that occurs with
high probability only near the location of a classical black hole.  The
virtual processes envisioned by Hawking and shown to be incompatible
with experiment in \bps , are suppressed by a small dimensionless
parameter.

Susskind has argued that this model does not produce world sheet physics
compatible with the existence of local observers, and in particular,
that it violates the world sheet cluster property.  His argument can be
phrased as follows.  Following the rules of string theory, the amplitude
for two events of ``black hole evaporation/ baby universe emission'' is
given by an integral
\eqn\babyamp{\int dx dy T(A(x) B(y))}
where both integrals are taken over the entire world sheet.  This is a
very nonlocal expression from the world sheet point of view.  In the
``conventional'' approach to wormhole physics, the superposition of
amplitudes with an arbitrary number of universe creation and
annihilation events is shown to be equivalent\foot{It is assumed that
the baby universes are always very small at the time of their connection
onto the rest of the world, and the equivalence referred to above
applies only to physics at a length scale larger than the typical baby
universe size.}to a superposition of
amplitudes for local field theories with different values of the
couplings.  Each of these separately satifies cluster decomposition.

Strominger instead proposes to enforce approximate world sheet locality
by insisting that world sheet physics is approximately classical and
that the vertex operators are nonvanishing only in the vicinity of
certain localized classical events.  However, the time ordering in
\babyamp appears to foil this attempt.  Consider two sequential black
hole formation processes separated by a large proper time in the
classical geometry.  Cluster decomposition requires that the $\$ $
matrix for this joint event approximately factorize into the $\$ $ matrices
for the single black holes.  The time ordering in the string theory
formula requires us to write the joint amplitude as a superposition of
two amplitudes corresponding to emission of the ``first'' baby universe from
either the ``first'' or ``second'' (according to the classical
evolution) black hole.  This enforces correlations on the joint dollar
matrix which are incompatible with factorization.  This ingenious attempt
to avoid the \bps constraints on information loss, appears to fail.\foot{
M.Srednicki ({\it Private Communication}) has pointed out that in the absence
of actual
experiments on evaporation black holes we have no right to invoke the
empirical success of the postulate of cluster decomposition.  Note also
that Polchinski and Strominger\ref\polstrom{A.Strominger, J.Polchinski,
hep-th 9407008.} have recently presented an
extension of Strominger's work and attempted to answer Susskind's objections.
I have been unable to understand their paper, but perhaps one should
not conclude prematurely that the idea of information loss to wormholes
is dead.}

\newsec{Horned Particles as the Endpoint of Hawking Evaporation}

The first intimations of a way out of the problems of the remnant scenario
appeared in an unpublished note by Dyson, very soon after Hawking's seminal
papers on the evaporation of black holes.   The interior geometry of a black
hole has, as we have seen, the structure of a collapsing horn or neck.  In the
case of a black hole formed in collapse, the collapsing matter sits down
at the very end of the horn.  Dyson envisaged a ``pinching off'' of the horn,
which would leave the collapsed matter in a separate closed universe, thus
``explaining'' the loss of information found by Hawking.

In fact, as noted above,
the ``pinching off'' of a closed universe cannot be adequately described
by the classical picture of Dyson.  A closed universe carries vanishing energy
and momentum, and by the uncertainty principle, the event of ``pinching
off'' can be localized neither in space nor in time.  Dyson's idea leads
instead to what Sidney Coleman has dubbed ``spacetime wormholes''\foot{Not to
be
confused with the spatial wormholes of Wheeler.}.

Although Dyson's picture is not directly relevant to the black hole information
problem, his idea that the information ``just goes somewhere else'' turns out
to be crucial.  Chronologically, the second step in the
unravelling of the mystery came with the work of Guth and Farhi
\ref\guth{A.Guth, E. Farhi, {\it Phys. Lett.}{\bf B183},(1987),149.}
on a subject they facetiously called ``the creation of a universe in the
laboratory''.
In modern versions of the inflationary universe, our current horizon volume
has inflated from a very small part of a much larger universe.
It is not necessary
for that larger space to have undergone inflation in order to create
the universe that we see.
Guth and Farhi asked whether a similar event could occur
somewhere in our own universe.  In other words, could a small region of the
universe find itself in a false vacuum, and if so, what would occur?
If one doesn't think
too hard about this problem it seems to pose a paradox.  The false vacuum
region has a higher energy density than the true vacuum and so the
interface between them is subject to an inward pressure.  Our flat
space instincts tell us that the false vacuum region should shrink.  On
the other hand, it has a positive cosmological constant, and therefore it
should grow.  Of course, dynamical curved space is able to accomodate
both of these features.  The false vacuum region evolves into an ``embolism''
(Fig. 5), a large inflating DeSitter space connected onto ordinary
Minkowski space by a rapidly shrinking neck.

\ifig\ffive{The Guth-Farhi Embolism: An Entire Universe Looks Like a Black
Hole Remnant to An External Observer}
{\epsfysize=6cm \epsfbox{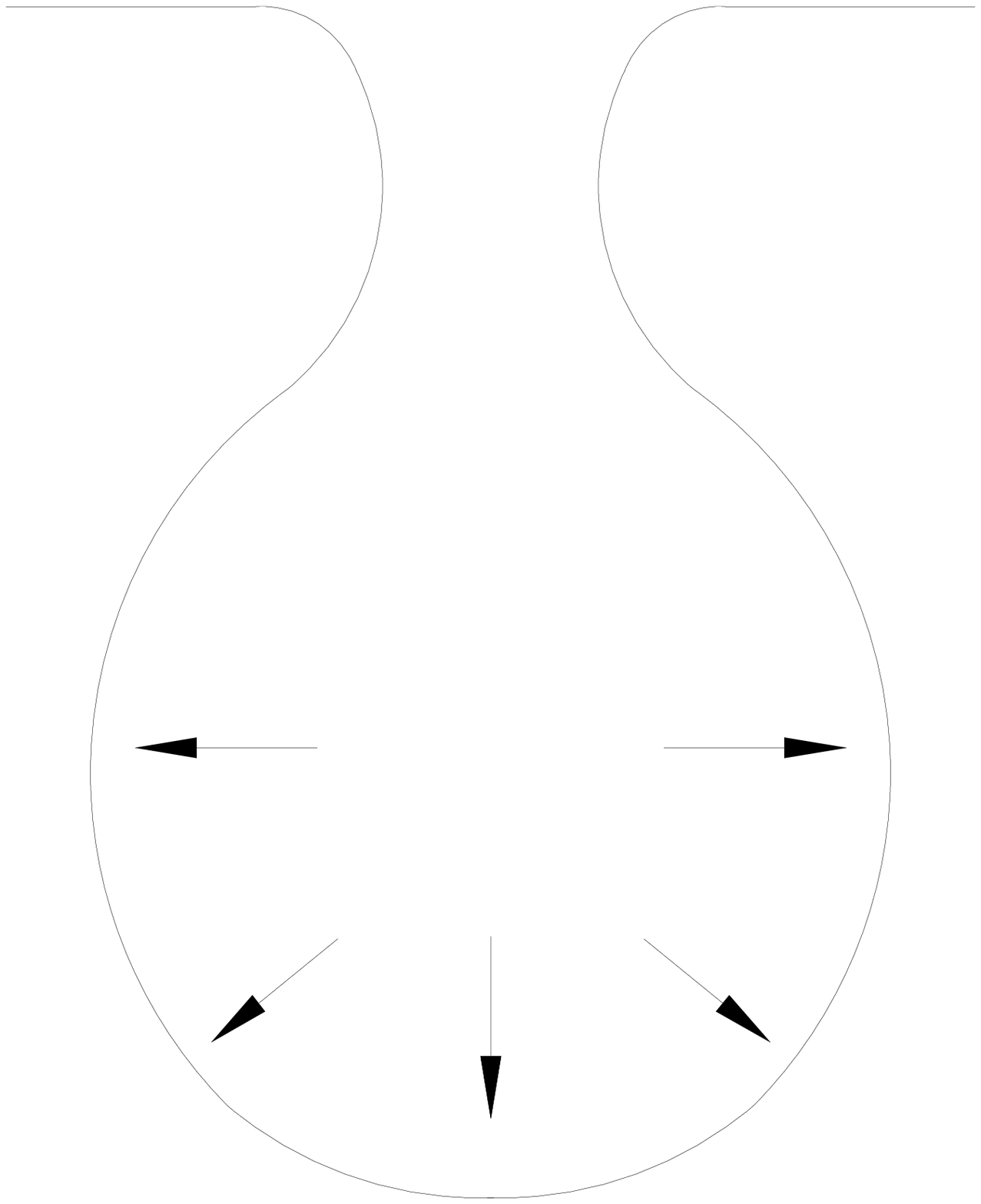}}

By Birkhoff's theorem
the geometry outside the neck looks like a black hole and we have an
example of a black hole with a large internal geometry\foot{Guth and Farhi
showed that the initial configuration with a localized region of false vacuum
could not evolve classically from nonsingular initial conditions,
but argued that it might arise from quantum tunneling}.  Subsequent evolution
is generally said to lead to ``pinching off'' of the neck, but all this really
means is that the neck evolves to a Schwarzchild singularity where
the equations of motion break
down and we cannot predict what really occurs
without the aid of a more elaborate theory of the short distance structure
of spacetime.

Suppose that the collapse of the neck is stabilized by some effect in this
mythical short distance theory.  The resulting object will look pointlike
to a flat space observer with limited resolving power.  He might be tempted
to call it a particle.  However, having been present at the creation, Guth
and Farhi tell us that it actually contains an entire universe, perhaps one
as large and as rich as our own.  Surely one must suspect that its behavior
might be quite different from that of an ordinary elementary particle,
particularly when
it comes to questions about whether it can be created from the
vacuum by the action of weak external fields.
This then is a type of geometrical object that seems quite reasonable from
the point of view of general relativity, especially if we are willing to
invoke (as yet ill understood) quantum mechanical or string theoretic
mechanisms to avoid the occurence of singularities.  The basic premise of
these lectures is that objects like the Guth-Farhi universe, are not just
a particular way to make a black hole, but are the natural endpoint of
evolution for all black holes.

We now come to the important question of a name for almost pointlike (from the
point of view of an external observer) black holes with large internal
geometries.  In the context of the physics of near extremal charged black
holes, where this circle of ideas arose, I proposed that they be called
{\it horned particles}, or {\it cornucopions}, ``to celebrate both the shape
of their internal geometry and the wealth of information hidden inside them''.
The name is, I think, appropriate for this general class of object,
because, as described in the introduction, the Schwarzchild geometry
itself is
a horned particle.  To see
this, we must dare to look behind the horizon. The
celebrated Schwarzchild singularity is (in internal Killing coordinates),
nothing but the squeezing down of the internal world into an infinitely long,
infinitely thin horn.  The ``horny'' aspects of the internal geometry of
a black hole are quite generic.
Even if we suppose that the transverse squeezing
is stabilized at some very small radius, there is no reason to suppose
that the stretching stops as well.  Indeed, we will see in the next section
that models of near extremal charged black hole collapse and evaporation,
lead to internal spaces which have large extent only in one dimension.

\newsec{Near Extremal Charged Black Holes}

The Hawking temperature
of Reissner-Nordstrom black holes in Einstein-Maxwell gravity,
vanishes in the extremal limit $Q = M$.  This is easily understood
in terms of the geometry of the extremal black hole.  For $Q=M$ the Killing
vector which is timelike at infinity is everywhere timelike, the singularity
is a timelike curve, and there is no horizon at any finite distance.
Although quantum field theory
on this background requires some kind of boundary condition on the singularity,
it has a time independent Hamiltonian.  Quantum fields
propagating on the background are in pure states\foot{I am ignoring
problems caused by the Cauchy horizon.  There is a region of the
spacetime, including the whole asymptotic region, which is causally
separated from the Cauchy horizon.}.

It has therefore seemed plausible to many researchers that extremal charged
black holes are the endpoint of Hawking evaporation for the case where a black
hole manages to retain its charge.  It must be emphasized that this may be
a very rare occurence in the real world.  For a charged black hole,
Schwinger pair production acts to drain off the charge with the Hawking
radiation.
Given the known spectrum of light charged particles it seems implausible
that a black hole could retain a very large electrical charge.
Magnetically charged black holes may have a better chance of retaining their
charge because magnetic monopoles are undoubtedly very heavy.  However,
in modern theories of elementary particle physics, monopoles arise as classical
solutions to the field equations, which may have internal variations on scales
much larger than the Planck length.  The Reissner Nordstrom solution for
 pointlike monopoles, might have little to do with these
objects\ref\nair{K.Lee, V.P.Nair, E.Weinberg, {\it Phys. Rev. Lett.}{\bf
68},(1992),1100, hep-th 912008.}
unless the magnetic charge is very large.  In that case the scale of
the Reissner Nordstrom geometry is much larger than that of the internal
structure of the monopole, and the magnetic field at the horizon is very
weak.  We might hope that such highly charged objects are quasi-stable.
The large (magnetic) charge limit is also the limit in which the
geometry becomes smoothly varying and we can hope to avoid the necessity
of understanding short distance physics in our discussion of Hawking
evaporation.

Most researchers have decided to ignore the possible real world
difficulties with the idea of stable charged black holes.  The charged
, near extremal black hole is viewed as a model in which the phenomenon of
horizon formation and evaporation can be studied within the framework of
a controlled approximation scheme.  One hopes to extract lessons from it
that may be applicable to more realistic but less tractable problems and
in
 particular to neutral black holes\foot{Recent work\ref\gidpostro{S.Giddings,
J.Polchinski, A.Strominger, {\it Phys. Rev.}{\bf D48},(1993),5748.} on
 exact
string theoretic classical solutions for the region of space inside the
 horn of a dilaton black hole seems to indicate the existence of
 extremal
{\it neutral } black holes, with transverse sizes of the order of the Planck
length.}. We will
follow this tradition, and say no more about the unrealistic features of
charged black holes.

The Reissner-Nordstrom solution is not lightlike geodesically complete.  Much
of the initial work in this area concentrated instead on the geodesically
complete charged black hole solution of the version of scalar-tensor (dilaton)
gravity
that arises in string theory\ref\GMGHS{G.Gibbons,R.Maeda,{\it Nucl. Phys.}
{\bf B298},(1988),741; D.Garfinkle, G.Horowitz, A.Strom-\break
inger, {\it Phys. Rev.}{\bf D43},(1991),3140.}.
Although this solution is also singular, it is geodesically complete and the
singularity is at infinity.  More importantly, in the appropriate conformal
frame, the singularity can be attributed to the blowup of the parameter which
controls quantum fluctuations, which suggests that quantum effects may well
cure the singularity.  Indeed, portions of
the geometry, shown in Fig. 1, are very similar
to the geometry
 encountered in $1+1$ dimension soluble string theories\ref\1dstr{See {\bf
The Large N Expansion in Quantum Field Theory and Statistical Physics},
E.Brezin, S.Wadia, Editors, World Scientific, Singapore, 1993, for a\break
comprehensive
collection of articles on one dimensional string theory.}
{}.
This geometry, whose form was the origin of the name cornucopion, has a large
region in which spacetime is approximately two dimensional.  The
solution \GMGHS
is, in that region, identical to the classical background of $1+1$ string
theory, in the limit of vanishing tachyon condensate $\mu$.  In string theory,
Knizhnik Polyakov Zamolodchikov (KPZ)\ref\kpz{D.Knizhnik, A.M.Polyakov,
A.B. Zamolodchikov, {\it Mod. Phys. Lett.}{\bf A3},(1988),819.} scaling tells
us that the
effective loop expansion parameter is $1\over \mu$.  The perturbation series
for the S matrix can be exactly computed and various resummations of it have
been proposed\ref\moore{G.Moore, M.R.Plesser, S.Ramgoolam, {\it Nucl. Phys.}
{\bf B377},(1992),143, hep-th 9111035.}.  All of them have the property that
the $\mu = 0$
{\it quantum} theory is well defined, despite the blowup of the terms in the
semiclassical expansion.  One may therefore hope that the singularity of the
GMGHS solution is similarly cured by quantum mechanics.

We emphasize again however that the whole focus of the work on charged black
holes is to find mechanisms that resolve the Hawking paradox.  One should not
put too much emphasis on a particular model.  Indeed, after the breakthrough
in understanding the dilaton gravity models,
Strominger and Trivedi reanalyzed the ordinary Reissner-Nordstrom black hole
and argued
that similar results about the existence of infinite numbers of degenerate
states and effective information loss could be established in that
context as well,
by concentrating on regions far from the singularity.  Indeed this model
may be argued to avoid the strong coupling problems of the GMGHS solution.

One final piece of justification before proceeding to an analysis of the
GMGHS solution:  in a realistic quantum string theory, we expect the dilaton
to acquire a mass, drastically changing the form of the black hole solution
in the asymptotic region.  However, unless the dilaton effective potential
blows up in the
horn of the cornucopion, it will not change the physics of the two dimensional
region on which we shall concentrate\ref\horowharv{J.Horne, G.Horowitz,
{\it Nucl. Phys.}{\bf B399},(1993),169, hep-th 9210012; R. Gregory, J.Harvey,
{\it Phys. Rev.}{\bf D47},(1993),2411, hep-th 9209070.}

\subsec{The CGHS Lagrangian}

In theories containing both a metric and scalar fields, the definition
of the metric is ambiguous because of the freedom to perform Brans-Dicke
transformations.  These scalar field dependent Weyl transformations can
change the structure of the lagrangian and the apparent form of
solutions to the field equations, but cannot change the result of any
physical measurement.  It is however often convenient to work in some
particular Brans-Dicke frame.

In string theory, the obvious choice of conformal frame is that dictated
by the string loop expansion.
This is computed in terms of a two dimensional
conformal field theory whose lagrangian has the form
\eqn\wslag{{\cal L}_{worldsheet} = G_{\mu\nu}(x) \partial x^{\mu}\bar
{\partial} x^{\nu} + \ldots}
The metric $G_{\mu\nu}$ which appears in this lagrangian is known as the
{\it string} metric or $\sigma\quad model$ metric.  At energies low compared to
the string scale, the lagrangian for the string metric, the electromagnetic
field and the dilaton takes the form:
\eqn\Cfmp{{\cal L} = e^{-2\Phi}\sqrt{-G}(R + 4(\nabla\Phi)^2 -{1\over 4}
F_{\mu\nu}F^{\mu\nu})}
Garfinkle Horowitz and Strominger\GMGHS found simple classical
solutions of this lagrangian which are conformal transforms of the
magnetically
charged dilaton black holes of Gibbons and Maeda\GMGHS.  In
the case of extremal charge, the solution takes the form
\eqn\cornmetric{ds^2 = -dt^2 + e^{4\Phi} d{\bf x}^2}
\eqn\corndil{e^{2\Phi}= e^{2\Phi_0} + {2Qe^{\Phi_0}\over |x|}}
\eqn\cornmag{F_{\mu\nu}dx^{\mu}dx^{\nu} = Q sin\theta d\theta d\phi}
Notice that the geometry, depicted in Fig. 1, is completely nonsingular and
has an infinite region which is approximately a cylinder with a two sphere
cross section.  The radius of the two sphere is of order $Q$.  For large
$Q$ the curvature is everywhere small.  The only singular aspect of the
solution is the blowup of the string coupling as we move down the horn
towards $x=0$.

Deep inside the horn, it is convenient to make a "Kaluza-Klein"
expansion of the fields in spherical harmonics.  It is easy to see that,
apart from the usual Kaluza Klein $SO(3)$ gauge field, all of the non
spherically
symmetric modes propagate as massive waves inside the horn.  Thus, there is
an effective two dimensional theory of long wavelength modes propagating
in the horn\ref\giddstrom{C.Callan, S.Giddings, J.Harvey, A.Strominger,
{\it Phys. Rev.}{\bf D45},(1992),1005; S.Gidd-\break ings, A.Strominger
hep-th 9202004.}
\eqn\twodlag{{\cal L}_{CGHS} = e^{-2\Phi}\sqrt{-G}(R + 4(\nabla\Phi)^2
+ {1\over Q^2}- {1\over 4}F^2 -{1\over 4}(G^a)^2 )}

Here all of the fields are two dimensional, $F$ is the two dimensional
electromagnetic field strength, and $G^a$ is the field strength of
the $SO(3)$ gauge field which serves as the carrier of information about three
dimensional spatial rotations in the effective field theory.

In the two dimensional region, the extremal GMGHS solution has
the simple form
\eqn\lin{G_{\mu\nu} = \eta_{\mu\nu}}
\eqn\lindil{\Phi = {x\over Q}}
which is called the linear dilaton vacuum.
The lagrangian \twodlag also has solutions representing nonextremal
black holes, which have the form
\eqn\blahole{G_{\mu\nu} = \eta_{\mu\nu}e^{2\Phi}= (1 - M x^+ x^-)}
Here, $M$ is the deviation of the four dimensional black hole mass from
extremality.  In the literature it is often called the mass of the two
dimensional black hole.
There are solutions corresponding to dual charged and rotating black
holes as well.

Dilaton black holes with large magnetic charge are certainly classical
solutions of heterotic string theory.  In that theory, or in any other
containing light electrically charged fermion fields, these solutions
will have a large number of fermion zero modes.  Far away from the
position of the black hole, these will be the Callan-Rubakov
\ref\callanrub{C.Callan, {\it Phys. Rev.}{\bf D26},(1982),2058; V.Rubakov,
{\it Nucl. Phys.}{\bf B203},(1982),311.}
modes of charged fermions around a magnetic monopole.
They fall into angular momentum multiplets with angular momentum $J = Q-1$.
(In most of the literature, they are mistakenly called S-wave modes.
They are only S-wave for minimal Dirac magnetic charge.)
Deep inside the horn of the black hole, these modes propagate as massless two
dimensional
fermion fields coupled to the gauge fields.  The effective lagrangian is
\eqn\fermlag{\bar{\psi_N} (i\hat{\partial} - e\delta_{NM} + J^a_{NM}
\hat{G}^a)\psi_M}
In the existing literature, the couplings to the two dimensional
electromagnetic
and Kaluza Klein gauge
fields have been neglected.  The fermions were treated as free apart from their
gravitational interactions, and were bosonized into a set of free scalar
fields, giving the
famous lagrangian of Callan, Giddings, Harvey and Strominger (CGHS)\ref\cghs{
C.Callan, S.Giddings, J.Harvey, A.Strominger, {\it Phys. Rev.}{\bf
D45},(1992),1005.}:
\eqn\CGHSlag{{\cal L} = \sqrt{|g|}\bigl{(}e^{-2\phi}[R + 4(\nabla\phi )^2 +
4\lambda^2 ] - \ha [(\nabla f_i
)^2 + \ldots ]\bigr{)}}
In fact, one can reproduce the results of CGHS in a way that justifies the
neglect of
the gauge fields.

The free fermion lagrangian is a conformal field theory.  It contains
operators,
$\bar{\psi}\gamma^{\mu}(1 \pm \gamma_3)\psi$ and $\bar{\psi}\gamma^{\mu}(1
\pm \gamma_3) J^a \psi$
which satisfy a $U(1)XU(1)XSU(2)XSU(2)$ Kac-Moody algebra, where the
nonabelian part has
level $2Q - 1$.   The full stress tensor of the theory can be written
\eqn\stress{T_{fermion} = T_{Sugawara} + T_{singlet}}
where $T_{singlet}$ commutes with all the Kac-Moody currents.  It has central
 charge which
is proportional to $Q$ for large $Q$.  The fundamental operators out of which
the singlet
stress tensor is constructed, are bleached parafermions, defined by
multiplying the original
fermions by approriate path ordered exponentials of the Kac-Moody currents.
They are not
local with respect to the fermion fields, nor even with respect to all of the
fermion
bilinears.  Nonetheless, the operator $T_{singlet}$ and the fermions are
relatively local.
Thus we can imagine constructing by local operations a state of the system in
which
$T_{singlet}$ has a classical expectation value, but the Kac Moody currents
all have expectation
value zero\foot{What is not at all clear is how to do this starting from the
full
four dimensional theory}.  Then, no gauge fields will be excited, only a two
dimensional
gravitational field.  The resulting classical equations in the gauge
$g_{\mu\nu} = \ha e^{2\rho}\eta_{\mu\nu}$ :
\def\pa{\partial_+}
\def\pab{\partial_-}
\eqn\classeqa{0 = {\delta S\over\delta\Phi} =
e^{-2\Phi} (4\pa\pab\rho + 8\pa\Phi\pab\Phi -8\pa\pab\Phi
+2\lambda^2 e^{2\rho})}
\eqn\classeqb{  0 = {\delta S\over\delta\rho} =
2e^{-2\Phi} (2\pa\pab\Phi - 4\pa\Phi\pab\Phi
-\lambda^2 e^{2\rho})}
\eqn\classeqc{0 = {\delta S\over\delta g^{++}} =
{T_{++}}_{singlet} +
e^{-2\Phi} (4\pa\Phi\pa\rho -2\pa^2 \Phi)    }
(with a similar equation for $\delta S \over \delta g^{--}$)
are identical with those of CGHS.  They have solutions corresponding to the
creation of
a black hole by an infalling pulse of leftmoving radiation.
However, the quantum expectation value of the trace of the stress tensor
$T_{singlet}$ is
large, of order $Q$, in any region where the curvature becomes of order 1.
Thus one is led
to construct a $1\over Q$ expansion in which this quantum correction, but no
other, is taken into
account self consistently.  A systematic expansion is possible when the black
hole
mass is of order $Q$ in the large Q limit.   The resulting CGHS equations are
\eqn\cghsa{ 0 =
e^{-2\Phi} (4\pa\pab\rho + 8\pa\Phi\pab\Phi -8\pa\pab\Phi
+2\lambda^2 e^{2\rho})}
\eqn\cghsb{   0 =
2e^{-2\Phi} (2\pa\pab\Phi - 4\pa\Phi\pab\Phi
-\lambda^2 e^{2\rho}) - 2\kappa \pa\pab\rho      }
\eqn\cghsc{0 =
 {T_{++}}_{singlet} +
e^{-2\Phi} (4\pa\Phi\pa\rho -2\pa^2\Phi)
- \kappa ((\pa\rho)^2-\pa^2\rho + t_+(x^+))   }
(and $++ \rightarrow --$), where the functions $t_{\pm}$ are determined by
the boundary conditions.
$\kappa = e^{2\Phi_0}Q$, is the usual rescaled coupling constant.
The nature of the solutions of these equations has been discussed extensively
in the literature
\ref\numwork{T.Piran, A.Strominger, hep-th 9304148; S.Hawking, I.Stewart,
hep-th 9207105;\break
 B.Birnir, S.Giddings, G.Horowitz, A.Strominger, {\it Phys. Rev.}{\bf
D46},(1992),638;
 J.Russo,\break L.Susskind, L.Thorlacius, {\it Phys. Lett.}{\bf B292},(1992),
13; L.Susskind, L.Thorlacius, {\it Nucl. Phys.}{\bf B382},(1992),123; D.Lowe,
{\it Phys. Rev.}{\bf D47},(1993),2446; T.Banks, A.Dabolkhar,\break
M.Douglas, M. O'Loughlin, {\it Phys. Rev.}{\bf D45},(1992),3607.}.
Here I will recall only a few salient points of that analysis.
The solutions do exhibit Hawking radiation, as well as the backreaction of
the geometry
to that radiation.  The apparent horizon (defined as the place where
$\partial_+ \phi = 0$)
recedes from the observer.  However, all of the solutions inevitably encounter
a
singularity.  The reason for this is apparent if we note that
\cghsb is singular when $e^{-2\Phi} = \kappa $.  Essentially this is
because, thinking of the system as a nonlinear model with $\Phi$ and
$\rho$ parametrizing the target space, the target space metric has zero
determinant at this point. Writing $\Phi = -\ha ln\kappa + \Delta$ and
taking linear combinations of the trace and dilaton equations near
$\Delta = 0$ we obtain
\eqn\trace{(e^{2\Delta} - 2)\pa\pab\rho = - 2 \pa\pab \Delta}
\eqn\dil{(e^{2\Delta} - 1)\pa\pab\rho = {3\over 2} (4\pa\Delta\pab\Delta
+ \lambda^2 e^{2\rho})}
Since , $e^{-2\rho}\pa\pab\rho \propto R$, the scalar curvature, we will
have a curvature singularity unless the right hand side of \dil vanishes
when $\Delta = 0$.  However, the expression on the right hand side
involves only first derivatives of the fields and will not vanish for
generic initial conditions.  In particular, if
the initial state of the system is the
linear dilaton vacuum, in which the field $\Phi$ attains the value $-\ha
ln\kappa$ at a finite point, we should expect generic perturbations of
it to produce a singularity.
The Penrose diagram for these singular solutions resembles that of an ordinary
black hole.

In attempting to resolve the singularity of the CGHS equations one might
imagine that it were
necessary
to resort to the full machinery of short distance quantum gravity or string
theory.
However, it is not clear that this is the case.  What is going wrong here is
that the coupling ,$e^{2\Phi}$, is getting large.  Our effective two
dimensional field
theory was derived from string theory in the classical approximation.
Perhaps the strong
coupling corrections to the effective lagrangian are large, but the degrees
of freedom
that we have isolated remain classical, and the evolution of the black hole
can still
be studied in terms of them.  We will explore this possibility in the next
section.

\subsec{Non Singular Lagrangians}

If we imagine integrating out the string theoretic and four dimensional degrees
of freedom that are neglected in the CGHS lagrangian, we should expect to
get a general renormalizable lagrangian for the graviton, dilaton, gauge
fields and
fermions.  The lagrangians that we will study are not the most general possible
lagrangians for these fields.  First of all, as before, we will not discuss
the gauge
interactions.  Furthermore, we will assume that the singlet conformal field
theory
with stress tensor $T_{singlet}$ is unchanged by higher order corrections.
There is
no real justification for this assumption, we make it only to keep the
system ``soluble''.
Up until this point our considerations could be justified (for large $Q$)
as a systematic approximation to a realistic model.  This will no longer be
the
case.  We are now studying a model of black hole formation and evaporation
which may not correspond to anything in the real world (even the real world
according to string theory).
Nonetheless, our model is a close cousin of a realistic one, and its
 general features could well persist in the more complicated systems
that we are unable to analyze.

The most general renormalizable lagrangian for the graviton dilaton system is
\eqn\genlag{{\cal L} = \sqrt{-g}(D(\Phi )R + F(\Phi )(\nabla\Phi )^2 - V(\Phi
))}
In the weak coupling limit $\Phi\rightarrow -\infty$, we have
\eqn\asympa{D\rightarrow e^{- 2\Phi}}
\eqn\asympb{F\rightarrow 4 e^{-2\Phi}}
\eqn\asympc{V\rightarrow  {1\over Q^2}e^{-2\Phi}}

The kinetic term in the lagrangian will be nonsingular for all $\Phi$ if $D'(
\Phi )\neq 0$.
If this is satisfied, the only singularities can come from the region of
field space $\Phi \rightarrow \infty$.
However, it is easily verified from the explicit solutions which we will
present below,that if
\eqn\asympaa{D\rightarrow e^{n\Phi}}
\eqn\asympba{W\rightarrow e^{m\Phi}}
with $n\geq m-2$ as $\Phi\rightarrow\infty$, then
the solutions are nonsingular there as well\ref\nonsing{T.Banks,
M.O'Loughlin, {\it Phys. Rev.}{\bf D48},(1993),698.}.
Here, $W$ is an auxiliary function which we will define below.

To solve these models it is convenient to do a field dependent Weyl
transformation $g_{\mu\nu} \rightarrow e^{2\Sigma (\Phi )} g_{\mu\nu}$,
which eliminates the $F$ term.  We obtain a lagrangian
\eqn\DW{{\cal L} = \sqrt{-g}(D(\Phi )R  + W(\Phi ))}
(with $W = e^{2\Sigma} V$).  Furthermore, since $D'(\Phi ) \neq 0$, we might
as well take $D$ to be the independent scalar field
in this lagrangian. We will use the notation $W(\Phi (D))\equiv W(D)$.  Note
that the equation of motion for $D$
involves no derivatives, so that this can be viewed purely as a gravitational
lagrangian with higher curvature interactions.
It is a general result that all solutions of all lagrangians of this type
have a Killing vector, and that $D$ is constant
along the Killing flows\ref\minkgrav{T. Banks, M. O'Loughlin,
 {\it Nucl. Phys.}{\bf B362},(1991),649.}.  Asymptotically, we want to tie on
to the
black hole solutions of the CGHS lagrangian and we thus take the Killing
vector to be timelike at infinity.
For this simple system, Killing horizons are precisely event horizons.
In Killing coordinates the event horizons are
thus at points where ${d D \over ds}=0$.

Our system is two dimensional, and we can thus find a Killing coordinate
system which is also
conformal $g_{\mu\nu} = \eta_{\mu\nu} e^{2\sigma (s)} $.
In this gauge a complete system of equations for the metric and dilaton is
\eqn\EOM{{d^2 D \over ds^2} = 2 {d\sigma \over ds}{d D \over ds}}
\eqn\EOMb{{d^2 D \over ds^2} = 2 e^{2\sigma}W}
Using the freedom to shift the conformal factor by a constant, the general
solution is given by quadratures as
\eqn\soln{e^{2\sigma} = {d D \over ds}=   \int^D W(x) dx}

The single physical integration constant is the constant of integration in
the integral of \soln .
This integration constant plays the role of the ADM mass.  It is important to
emphasize that the linear dilaton vacuum
is not a solution of these equations.  This is as it should be.  The equations
we have written are supposed to reflect
the effective result of integrating out short distance quantum fluctuations.
We
should only expect their solutions to resemble the classical solutions of the
theory
in the extreme weak coupling region.  It is not even correct to identify the
ADM mass as the coefficient
of the leading correction to the linear dilaton behavior at infinity.
What we should do instead is to find a truly static (global timelike Killing
vector) solution of the equations
and define the ADM mass to be extremal at that value of
the integration constant for which the static solution is realized.
Nonextremal ADM mass corresponds to deviation from this static solution.

The nature of the solutions clearly depends crucially on the behavior of the
function $U(D)\equiv \int_{D_0}^D W(x) dx + M$
where $D_0$ is chosen so that $M=0$ is the static solution (if any).  If
$U(D)$ has precisely one real isolated zero for all values of $M$,
the geometry has a single horizon, which is always a finite spacelike distance
away from
finite points in the external region.  To see this, note that the distance is
given by
\eqn\dist{distance = \int ds e^{\sigma} = \int {dx \over \sqrt{U(x)}}}
There is no natural endpoint for Hawking radiation in such models.  We may
expect that when coupled to matter
fields they will lead to runaway solutions which Hawking radiate forever
\ref\runaways{A.Bilal, C.Callan, {\it Nucl. Phys.}{\bf B394},(1993),7.;
S.P. deAlwis, {\it Phys. Lett.}{\bf B289},\break (1992),278, {\it Phys. Lett.}
{\bf B300},(1993), 330;
S.Giddings, A.Strominger, {\it Phys. Rev.}{\bf D46},\break (1993),2454.}.
In \nonsing models with two, generically isolated, zeroes  were studied.
These have solutions with the causal
structure of Reissner-Nordstrom black holes, but without the timelike
singularities of the Reissner-Nordstrom
solution.  In particular, there is an extremal value for M at which the two
zeroes coincide.  In this case, the horizon
recedes to infinite spacelike distance, as one can see from \dist.  The
resulting metric is truly static, and quantum fields
placed in it will not Hawking radiate.  However, it is not lightlike
geodesically complete, and like the extremal RN
solution, it has a Cauchy horizon.

Martin O'Loughlin has recently invented another class of lagrangian, which
has a more satisfactory candidate for a
remnant solution.  In \nonsing no fine tuning of parameters was allowed.
The idea was to find generic behaviors of
large classes of lagrangians.  In this way it was hoped that one could
mitigate the fact that we do not know
how to calculate the functions $D,F$ and $V$ in realistic examples.
O'Loughlin's analysis applies to a large class
of lagrangians but requires one fine tuning of coefficients.
Returning to the case where $U(D)$ has only one real
zero for all M, he tunes a parameter in the lagrangian so that for $M=0$
(by convention), a pair of complex
zeroes merges with the real zero, making it into a triple zero.  Then,  the
spacelike geodesic distance to the horizon becomes
infinite in the extremal limit, and with the proper asymptotic behavior of
$W(D)$ as $D\rightarrow 0$, we can make the interior of the black hole
asymptotically
DeSitter.  The Penrose diagram of the black holes in this model is that of
Fig.2.  The distance to the horizon
is finite in the nonextremal case and infinite for $M=0$.

In order to study Hawking evaporation of these nonsingular black holes, we
must couple them to matter and compute the
backreaction in some kind of large N approximation.  When the back reaction
term is added to the equations, we must
impose a stronger condition $|D'(\Phi )| > \kappa$, where $\kappa$ is the
rescaled large $N$ coupling, in order to
avoid a CGHS type singularity.  With this constraint imposed, the equations
were
studied numerically for the case of
$U(D)$ with two zeroes, by O'Loughlin and Lowe\ref\olow{M.O'Loughlin, D.Lowe,
{\it Phys. Rev.}{\bf D48},(1993),3735.}.  As mentioned, this case is very
similar to the Reissner-Nordstrom
geometry and \olow also carried out a numerical study of evaporation of
Reissner-Nordstrom black holes.  This case
had been treated previously by Strominger and Trivedi
\ref\sandip{A.Strominger, S.Trivedi, {\it Phys. Rev.}{\bf D48},(1993),5778.},
using approximate analytical techniques.
The evolution is completely nonsingular\foot{There is of course a timelike
singularity in the Reissner-Nordstrom case, but this
does not effect the qualitative nature of the evolution}, and leads to the
following qualitative picture (Fig. 6).

\ifig\fsix{Evolution of the Spatial Geometry of a Nonextremal Black Hole
Back to an Extremal Remnant}
{\epsfysize=6cm \epsfbox{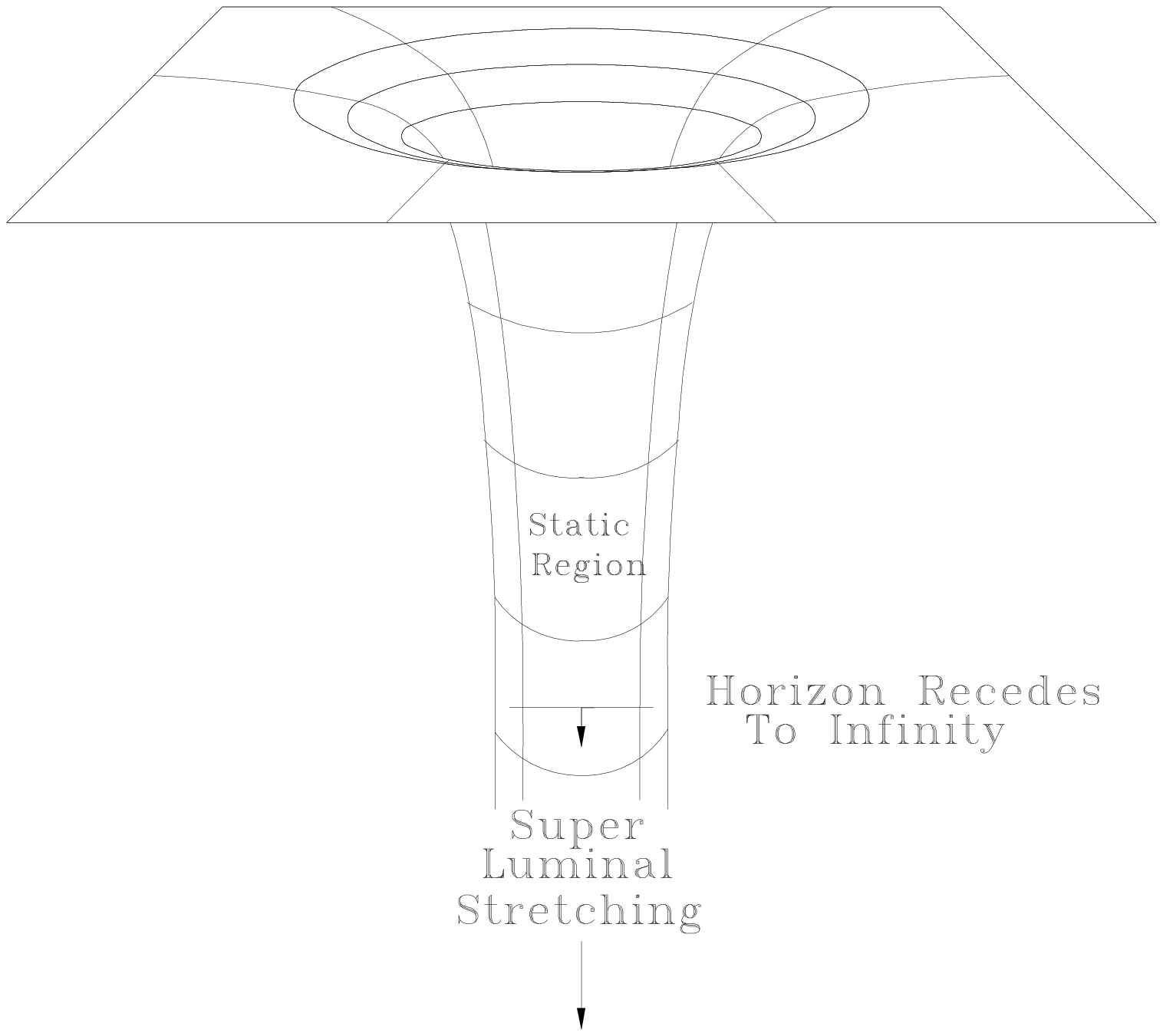}}

When matter is incident on an extremal black hole, an apparent horizon is
formed and the black hole begins to radiate.
The apparent horizon recedes from the external observer, eventually leaving
an infinite static geometry identical to the exterior
of the extremal black hole solution.  The full spacetime however has a
horizon, and one can verify explicitly that the
state of the field theory behind the horizon depends on the nature of the
initial infalling matter.
Thus, within these models, we have a consistent remnant scenario for black
hole evaporation,
in which information is lost to the external observer.  There is a different
type of remnant for each
kind of initial pulse that forms the black hole (and thus an infinite number
of different remnants
altogether), but they are all indistinguishable from the point of view of the
external observer.  The information
that distinguishes between them is causally disconnected from him.  On any
finite time slice, the system
still has an apparent horizon that is accessible to the external observer,
and it may be possible for him
to associate the information with states on the horizon.  Asymptotically
he cannot do so.  The apparent horizon
has gone off to infinity, a different infinity than his own initial
asymptotic region.  His S-matrix will not be unitary.

One issue that has not been sorted out by these studies is the fate of the
Cauchy horizon of
the initial black hole.  More properly, one should ask whether the true
spacetime geometry, the
solution of the large N equations, has a Cauchy horizon.  All that is known
for sure from
the numerical work is that no singularity was encountered in the interior of
the horizon within the time span covered by the calculation.  Possible causal
structures
that could be attained via nonsingular evolution, which would not
contain Cauchy horizons, have the form of Fig. 7
(Fig. 2 is a special case of this).

\ifig\fseven{Possible Causal Structures of Black Hole Remnants}
{\epsfysize=6cm \epsfbox{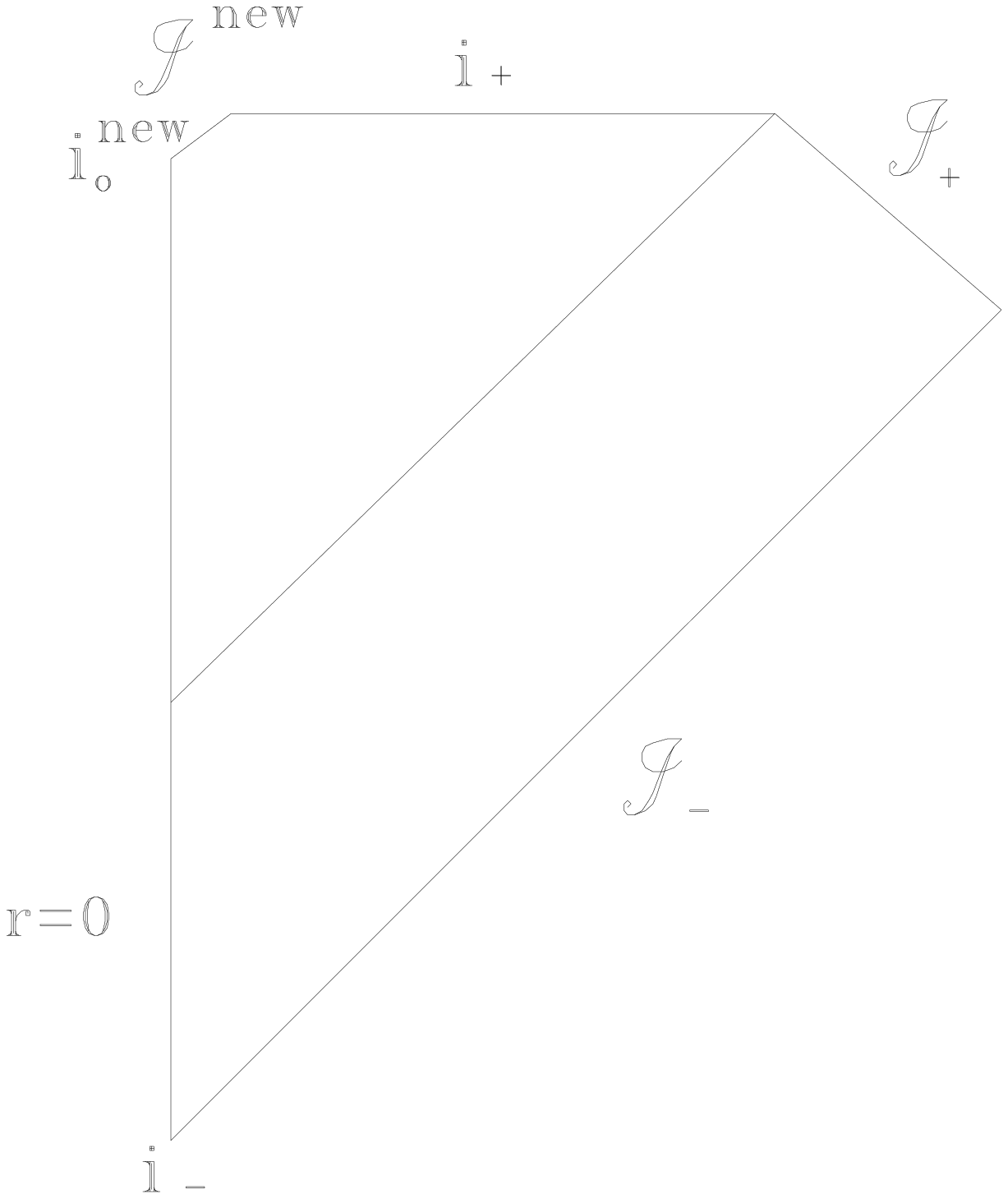}}

That is to say, the structure would be similar to that of the classical
solution of O' Loughlin's triple
zero model.  It would be of interest to resolve this issue and/or to study
the large $N$ evolution of
O' Loughlin's model.  With regard to the latter project, one should take
note of the probability that once quantum corrections are taken into account,
we will
have to retune the parameters of the lagrangian to have a horizon that is an
infinite
distance away.

\subsec{Production of Remnants from the Vacuum}

I now come to the discussion of the famous ``problem'' of the remnant
scenario, the spectre of
infinite production cross sections.  In doing so I must confess to a gross
error that I have made
in much of my published work on this subject, which has misled others as well
as myself.  In early
work\ref\corn{T.Banks, A.Dabolkhar, M.Douglas, M. O'Loughlin, {\it Phys. Rev.}
{\bf D45},(1992),3607.} I stated that the infinite numbers of degenerate
states of the cornucopion
were in fact causally connected to the external world.  They were supposed to
be stationary quantum
excitations of the static geometry located far down the horn of the
cornucopion.  Their degeneracy
was attributed to the fact that they were infinitely far away.  In fact, as
we have seen, the picture
is quite different.  On any finite time slice in a synchronous gauge,
a cornucopion is {\it not} a static geometry.  It has an apparent horizon,
behind which the spatial
geometry is undergoing superluminal expansion.  The infinite number of
degenerate states of the
cornucopion are the states of quantum fields lying behind this horizon.

The model for all discussions of production of nontrivial classical field
configurations from
the vacuum of a quantum field theory, is a classic paper of
Affleck and Manton\ref\affmant{I.Affleck, N.Manton, {\it Nucl. Phys.}
{\bf B194},(1982),38.}.  These authors studied pair production
of `t Hooft-Polyakov monopoles in a constant magnetic field, and showed
that in the small field limit the amplitude reduced to Schwinger's formula
for the corresponding amplitude for elementary monopoles.
The essential idea of the calculation is that for sufficiently small
accelerations, the only degree of freedom of a soliton that one should be
able to excite is the noncompact collective coordinate for its motion through
space.  Thus, for the motion of of a soliton in a weak external field, one
can construct
an approximate solution in which the soliton simply follows a curved world
line.
The idea of Affleck and Manton was that one can compute the pair production
cross section for solitons in the same field by analytically continuing
the solution for a pair of particles to Euclidean time.  The uniformly
accelerated hyperbolic Minkowski motion continues to a circle in Euclidean
time and one computes the production amplitude as the exponential of the
actionof
this Euclidean instanton.

We do not really understand the Euclidean continuation of quantum gravity, but
at the semiclassical level, there is a long tradition of seemingly sensible
calculations\ref\eucgrav{S.Hawking, G.Gibbons, {\it Phys. Rev.}
{\bf D15},(1977),2752; D.Gross, M.Perry, L.Yaffe, {\it Phys. Rev.}{\bf D25},
(1982),330;
 S.Coleman, F.Delucia, {\it Phys. Rev.}{\bf D21},(1980),2133.} which simply
take over the idea of analytically continuing
Minkowski geometries.  Special provisions must be made for situations in which
the vector field which we use to define time at infinity, changes signature
in some finite
portion of the spacetime.  The analytic continuation is performed by taking
this
vector field to be imaginary, but this only produces an Euclidean manifold in
the region
where the vector field is timelike. On the surface on which the change takes
place
the vector field is null.  In the analytically continued spacetime,
this null surface is replaced by a point, and the portion of the Minkowski
manifold beyond
this null surface is discarded.
Now consider the pair production of the degenerate states of cornucopions.
Unlike solitons,
these are not globally static configurations. They have, to a good
approximation, a
timelike Killing vector outside the apparent horizon\foot{This approximation
becomes
better and better as time goes on and the horizon recedes.}, but it becomes
null
on the apparent horizon.  Following the rules of Euclidean gravity as we know
them,
the Euclidean continuation of a cornucopion trajectory does not contain the
portion of space where the degenerate states of the cornucopion live.
Thus, the tunneling process cannot produce these states.

Although this is the basic argument against infinite production cross sections
advanced
in \ref\bos{T.Banks, M.O'Loughlin, A.Strominger, {\it Phys. Rev.}{\bf D47},
(1993),4476.}, that paper actually addressed a slightly different question,
and was motivated by the mistaken idea that the degenerate states were causally
connected to the external world.  The issue was whether the infinite static
geometry of the cornucopion could be produced from the vacuum by weak external
fields.
As we have seen, this is not really the correct question to ask if we are
worried about the
degenerate remnant issue, but it is nonetheless an interesting question.
The intuitive idea behind the argument of \bos was that unlike a t Hooft
Polyakov monopole,
the internal structure of a cornucopion was unstable to small perturbations.
Classically,
this is related to the Second Law of Black Hole dynamics.  Any small
perturbation
of the extremal black hole should produce a horizon.  In \bos it was argued
that
this was true in the case of perturbations that tried to move the center of
mass
of the cornucopion.  If this is so, then analytic continuation of the moving
solution
will lead to an instanton which creates only finite spatial volumes, namely
only the volume of space in front of the horizon of the moving black hole.
This had been shown previously for extremal Reissner Nordstrom black holes by
Garfinkle and Strominger\ref\garfstrom{D.Garfinkle, A.Strominger,
{\it Phys. Lett.}{\bf B256},(1991),146.}.  Recently, this claim has been
challenged in the
case of dilaton black holes in string theory
\ref\horgidd{F.Dowker, J.Gauntlett, S.Giddings, G.Horowitz, {\it Phys. Rev.}
{\bf D49},(1994),2909,hep-th 9312172.},
but it is unclear whether the infinite instantons that have been exhibited
really contribute to production processes.  The quantum corrections around
them are not well
defined.  One should also be wary of studying the process in a truly
constant background field.  The resulting spacetimes are not asymptotically
flat and it may be that one can get confused about what states are really
being produced.  This issue deserves further study, because it may
have a bearing on the single aspect of this entire subject that might be
amenable
to observational verification, the discovery of cornucopions.

\subsec{Remnants of the Imagination?}

If black hole remnants exist, will we ever be able to find them?  And if we do,
will we ever be able to tell what they are?  Unfortunately, the answer to the
first question is probably no.  Cornucopions are by hypothesis stable.
Although we have no very good idea about the processes which might produce them
in the very early universe, it is unlikely that the production process is
so finely tuned that it can produce a density of remnants that is
neither much larger nor much smaller than the density of ordinary matter
in the universe.  In the former case, cornucopions would be ruled out
by observational astronomy, in the latter we would never be able to find them.
We can save a model of the production process that produces too many
cornucopions by invoking inflation, but in that case there are likely to be no
black hole remnants in our portion of the universe.  We would have to wait for
macroscopic black holes to finish their Hawking evaporation before we could
get our
hands on real remnants.

But suppose we did so.  Would we then be able to tell what the remnants were
by their properties, or would they just behave like ordinary elementary
particles?  Classically I believe the answer to this question is no.
Small perturbations of the cornucopion change its internal structure and
produce
horizons\foot{Though this is where the issues raised by \horgidd must be
faced.}.
If we now imagine quantizing the classical cornucopion solution, it would
formally have
statistics, like any other elementary particle.  However, any realistic
experiment in which we attempted to measure these statistics would be doomed to
failure.  The procedure of scattering the cornucopions to measure their
statistics would
invariably cause the formation of an internal horizon and the emission of
Hawking radiation that would change the internal state in an uncontrollable
way.
The statistical phase would be unmeasurable, and the cornucopions would behave
like
classical distinguishable particles.  A double slit experiment for
cornucopions would resemble Fig. 8.

\ifig\feight{A Double Slit Experiment for Cornucopions}
{\epsfysize=6cm \epsfbox{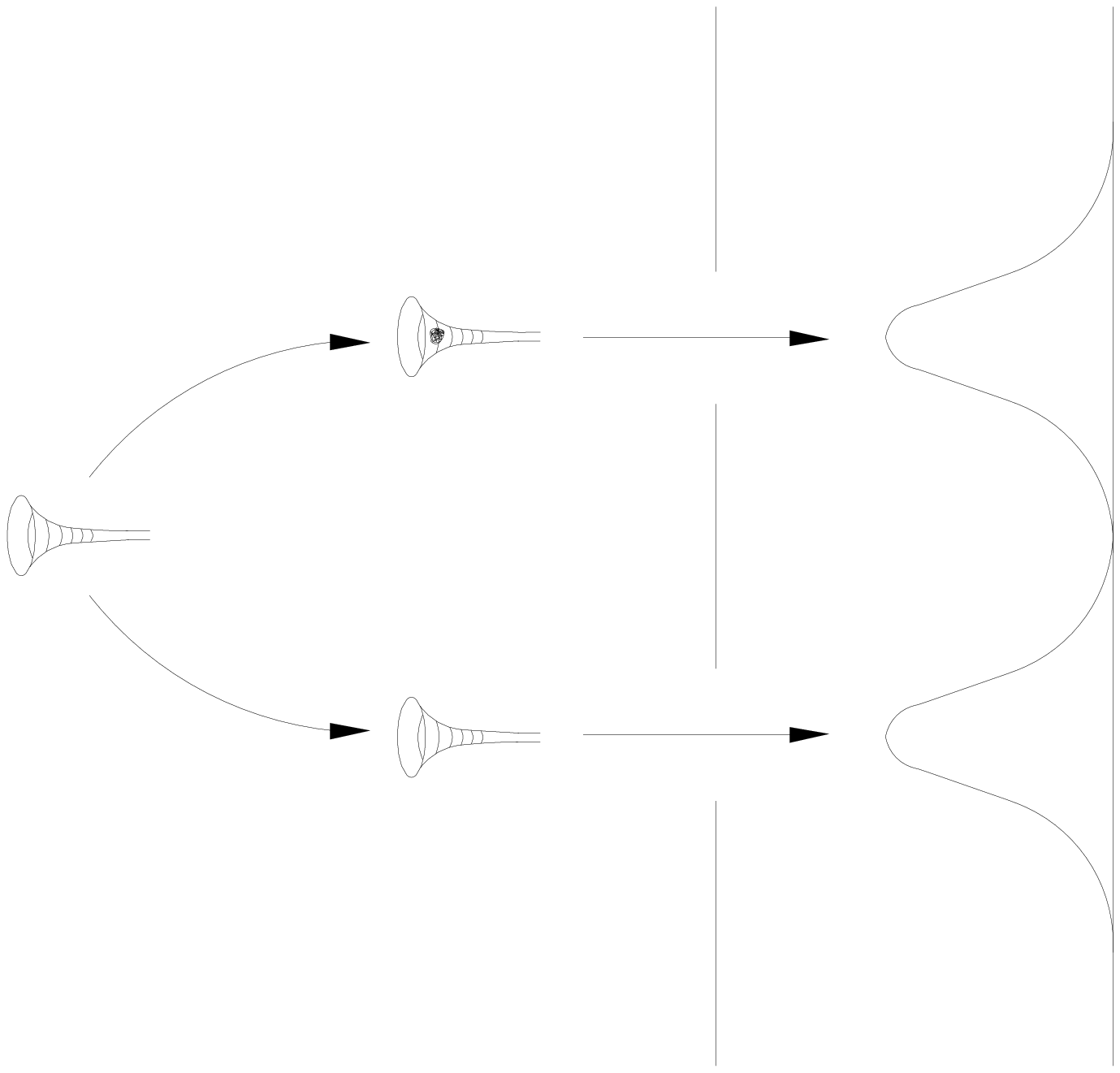}}

Apart from the objections of \horgidd , there is another way in which the
above argument
could fail.  Quantum mechanically there might be a threshold energy for
horizon formation,
as there is in the two dimensional model of Russo, Susskind and Thorlacius.
Given their infinite internal geometry, it is inevitable that the spectrum of
excitations
of cornucopions will have a continuum, but the continuum might be
separated from the ground state by a gap.  Experiments in which the
cornucopions were
moved with sufficiently small velocity would not excite the continuum
and in these experiments they would behave like quantum particles.
Determination of the size of the gap in the cornucopion excitation spectrum
seems to me to be one of the most interesting open problems in the study
of black hole remnants.  Unfortunately, it is a very hard problem, and
probably requires
the study of strongly coupled string theory.

Finally one should note that the pessimistic assessment of the probability of
finding black hole remnants
in the first paragraph of this section might be wrong.
Perhaps the differences between them and elementary particles
are sufficiently great that their production probability in an inflationary
universe
is much larger than we presently imagine.

\newsec{Unitarity, Information Loss and CPT}

The general picture that arises from the discussion of the previous
section is that black hole collapse and evaporation leads to the
formation of a new asymptotic region of space which is causally
disconnected from the old asymptotic region.  The spacetimes that we
have discussed can be foliated by spacelike hypersurfaces
 and within the semiclassical approximation for the
geometry, there will be no breakdown of unitarity.  Local physics will
continue to obey the rules of quantum mechanics.

It is clear however that the S-matrix measured by observers in the
original asymptotic region will not be unitary.  The cornucopion
spacetime has a horizon buried deep within the horn of the cornucopion.
On any spacelike slice, the state of the system
outside the horizon, which is all that can ever be probed by the
original asymptotic observer, is correlated with that inside the
horizon.  The external density matrix is not pure, and the purely
external scattering process must be described by a Hawking $\$$ matrix.

The cornucopion scenario thus unifies the idea of information loss in
the observable universe with the idea of black hole remnants, retaining
the merits of both while discarding their difficulties.  In particular,
Hawking's original claim that formation and evaporation of ``small
virtual black holes'' would lead to a reformulation of the fundamental
microphysical laws to accomodate time dependent information loss is
discarded, and with it the problems unveiled in \bps .  Virtual quantum
fluctuations of geometry that do not change the topological properties
of space, will be describable as distortions of the classical background
over small volumes, which subside after a short time.  No new asymptotic
regions are created in such fluctuations, and it should be possible to
``integrate them out'' and construct a local effective lagrangian for
long distance physics.  This local Lagrangian will obey the usual rules
of quantum mechanics.  Small, topology changing processes can also be
integrated out \worm , and they too lead to a picture in which the information
content of a single connected component of the universe does not change
with time.  Thus, to paraphrase J.A.Wheeler, we have {\it Information
Loss Without Information Loss}.  The rules of quantum mechanics are
locally preserved but the global S-matrix will not be unitary unless we
take into account contributions from causally disconnected asymptotic
regions.

Hawking has emphasized that any remnant scenario for black hole
evaporation implies a violation
of CPT in the observable universe.  This is indeed the case.
Suppose that it is possible to scatter a pair of elementary particles with
sufficiently
large energy and momentum transfer that they form a black hole.  The
resulting hole will Hawking radiate, and asymptotically settle down into
a cornucopion state in which a new asymptotic region of space is formed.
The CPT inverse of this process involves a conspiracy of the matter in
this infinite asymptotic region, which causes the ``internal universe''
to collapse and spew out precisely two particles into the region of
space in which we live.  On the other side of the horizon in the
meantime, particles must be sent in from infinity to converge on the point
where the cornucopion throat is sitting, and be absorbed by it as
inverse Hawking radiation.
Such initial conditions, involving as they do a rather special
conspiracy between causally separated points in two infinite universes,
are clearly of measure zero in the space of all possible initial
conditions.  CPT is violated in the same way that it
usually is in macroscopic systems: the inverses of processes which
increase coarse grained entropy require very special initial conditions,
which can never be realized in practice.
Once cornucopions are formed, there is essentially
zero probability that they will spontaneously dissipate.  Similarly, there is
zero probability that two cornucopions of opposite charge will annihilate
when the mouths of their horns meet.  Even if their charges
are opposite, the probability that their internal states
are exactly CPT conjugates of each other is zero.  This is not just a question
of the huge number of possible degenerate states available in the large
interior
of the cornucopion.  The interior state of the cornucopion is time dependent,
and involves an expanding geometry.  The CPT conjugate state is one with a
contracting geometry and very special initial conditions.  It will not
be realized in a cornucopion which formed from the collapse of matter in the
external world.

Indeed the annihilation process for cornucopions should be the time reverse of
the production process that we discussed above.  That process, as we saw
above, was both very improbable, and resulted in production of cornucopions of
very small volume.   Thus, in order to annihilate large volume, expanding,
cornucopions, we must first force them to evolve classically into the
configuration  which is produced at the end of the tunneling process described
in \bos .  This is very unlikely to happen.  Furthermore, causality prevents
it from happening as a consequence of simply moving the mouths of the
cornucopions so that they coincide in the external space.
Cornucopion production, whether through the improbable process of
pair production in an
external field followed by classical expansion of the internal
geometry, or through gravitational collapse, is an effectively irreversible
process from the point of view of a single causally connected sector of
the universe.

In passing, we note that
the above discussion is the strongest argument in favor of the existence of
neutral black hole remnants.  We can certainly imagine bringing together the
mouths of two oppositely charged cornucopions, obtaining a neutral object.
The above arguments suggest that that neutral object will have a large and
complicated internal geometry, which will not disappear.  It will be a neutral
cornucopion.  This is even more apparent if we imagine neutralizing the
charge of a cornucopion by dropping an elementary particle of opposite
charge into it.  This microscopic perturbation cannot destroy the
complicated internal geometry.  It is perhaps not surprising then that
Giddings, Polchinski, and Strominger have found what appears to be the
infinite horn of a {\it neutral} cornucopion as an exact classical
solution of string theory\gidpostro .

\subsec{Entropy}

No discussion of information loss and unitarity in black hole physics is
complete without
mention of the Bekenstein-Hawking entropy formula.  It is common to
describe this
entropy as a measure of the number of internal states of a black hole.
In the
context
of the cornucopion scenario this cannot be the case.  In this picture, a
black hole has an
enormous internal geometry and an essentially infinite number of
internal states.

A black hole of
mass M could be in any one of an infinite number of states depending on
its past history.  The information theoretic entropy of the external
world would depend on which state the hole was in.  However, most of the
information contained in the correlations between the internal and
external states of the black hole can have no impact on the future
interaction of the black hole with the external world, because it is
causally disconnected from the external portion of spacetime.
  Thus, this measure of entropy can
have no bearing on the {\it thermodynamic} entropy of the hole, which
describes how it exchanges energy with the outside world, and by
definition can depend only on our choice of spacelike surface in the
exterior.

Indeed, imagine an experimenter on Earth preparing a large number of
Einstein Rosen Podolsky pairs of neutrinos, and sending
one particle of each pair off to a distant galaxy.  The local state of the
Earth is thus rendered highly impure and incoherent.  However, the state
left behind by the departing neutrinos does not resemble that
produced by a black body sitting a finite distance from the earth.
The process of
pumping energy into the system of escaping neutrinos
(by sending more neutrinos out) will not be describable by the
laws of thermodynamics.  The neutrinos do not interact with the Earth
once they have been emitted.  A system can come into thermal equilibrium
only with a collection of states with which it is in causal contact, and
constantly exchanging energy.

The thermodynamic entropy of a large black hole can only be related to
states located very close to the horizon.  If we look at states
localized at any finite distance from the horizon, with a resolution
coarse enough so that the semiclassical picture of the geometry is
valid, then these states are expanding away from the horizon at
superluminal velocities.  They are out of causal contact with the
outside world, and in the cornucopion scenario they will remain so
forever.  It is only states ``infinitesimally close'' to the horizon
that can be in causal contact with the outside.  Here infinitesimally
close probably means within a length scale $l$ which is small
enough so that experiments probing physics
on the scale $l$ cannot be described by the semiclassical approximation.
This probably means that $l$ is of order the Planck length, or the slightly
larger
fundamental length of string theory.  It is thus
plausible that the number of states of a black hole that might be in
causal contact with the external world is proportional to the area of
the horizon, and it is perhaps natural that the proper units (i.e. those
in which the proportionality constant is of order one) of area are
Planck units\foot{In string theory, the string length is the more
natural unit.  This discrepancy will have important consequences below.}.

The fact that the information theoretic entropy of a black hole must, in
the remnant scenario, be thought of either as infinite (the logarithm of
the number of
possible final states of the remnant) or as dependent on the black
hole's entire previous history (the entropy of entanglement of the
internal state produced in a particular process of formation of the hole
with the external world) has been among the many arguments levelled
against the remnant scenario.  I believe that the above paragraph shows
clearly that neither of these quantities is a relevant measure of the
thermodynamic entropy of the hole.  Further it is clear that if the
black hole has a thermodynamic entropy, it should be proportional to the
area of its horizon.  A microphysical demonstration that the vicinity of
the horizon really contains the number of states indicated by the
Bekenstein- Hawking entropy formula would seem to depend on knowledge of
physics at very short distances.  Here we make contact with the point of
view of 't Hooft\ref\'tH{G. 't Hooft, {\it Nucl. Phys.}{\bf B256},(1985),727.}
 who has long insisted that the divergence of
the entropy of entanglement of the state of a quantum field theory
outside the horizon with that of the inside was the key to understanding
the BH entropy formula.  't Hooft believes that this demonstrates that
the understanding of black hole entropy and of the information paradox
is a problem of short distance physics and will guide us in the
construction of the fundamental theory of small scale geometry.  While I
do not agree with his assertion that unitarity of the S-matrix for the
original asymptotic region is a necessary ingredient in the construction
of the theory, I do agree that the thermodynamic nature of black holes
can only be understood in terms of short distance physics.

I have emphasized throughout this
review that our discussion of the evaporation of black holes
 depends only minimally on Planck scale physics and
not at all on quantum gravity.  We have seen that it is possible
to make the idea of information loss to an asymptotic observer
consistent with the basic rules of unitary quantum field theory in
a self consistently generated classical gravitational field.
The microphysical derivation of the
Bekenstein Hawking formula is the one aspect of this subject that still
seems to hint at the need for a more fundamental theory.  I believe that
I have outlined a way in which this formula could be compatible with the
idea of remnants, but the derivation of the formula itself seems to be
outside the domain of reliability of the semiclassical field theoretic
considerations that have been our guide up to this point.
Furthermore, this discussion makes it clear that the remnant scenario
(nor, I believe any scenario which leads to effective information loss)
cannot account for the information corresponding to the BH entropy.
In the next section I will describe one attempt that has been made to
understand the Bekenstein Hawking entropy formula from a microscopic
point of view.

\newsec{Stranded Strings and Black Hole Entropy}

The cornucopion scenario for black hole evaporation is a logically
complete, (though technically incomplete) description of this process.
By contrast, the material I am about to discuss is very much ``work in
progress''.  Its authors (primarily 't Hooft, Susskind, and their
collaborators), do not pretend to have achieved a logically complete
understanding of the process of Hawking
evaporation.  I am in the unfortunate situation of
understanding even what these authors {\it have} accomplished only
imperfectly.  I enter into a public discussion of their work with great
trepidation, and apologize in advance for my inevitable distortion of
their points of view.  All errors in the paragraphs that follow are my
own.

One of the fundamental ingredients in 't Hooft's work on black holes is the
Bekenstein-Hawking entropy formula.  One starts from the
assumption that black holes behave like thermal objects with Hawking
temperature and Bekenstein-Hawking entropy, and attempts to make this
fact consistent with the postulate that the S-matrix measured by the
initial asymptotic observer is unitary.  A special role for Schwarzchild
coordinates is implicit in these assumptions.  The black hole is static,
and thermodynamic considerations make sense, only in these coordinates.

One of the most revealing of the calculations performed by 't Hooft, is
that of the entropy of states of quantum fields in a small region of a
fixed Schwarzchild time slice, very near the horizon.  't Hooft argues
that in order for a quantum field theoretic calculation of the entropy
and energy of the black hole to give an answer of the order of the
classical values (i.e. $M$ for the energy and the BH entropy), one must
cutoff the quantum field theory by insisting that there are no states
at a proper distance closer to the horizon than $\sqrt{N \over 90\pi}M_P^{-1}$
where $N$ is the number of field degrees of freedom.  But this removes
many of the modes of the field which are involved in Hawking's calculation
of the outgoing density matrix.  Modes which can escape to infinity
with energies of order the Hawking temperature, have extremely high energy
near the horizon and do not satisfy 't Hooft's brick wall boundary
condition.

Susskind's point of departure is the 't Hooft entropy calculation
described above.  He brings in ideas from the Membrane paradigm, which
has had many successful applications to the astrophysics of black holes.
The basic idea of the membrane paradigm is that, from the point of view
of an external observer, things that fall through the horizon of a black
hole may equally well be imagined to reside on a membrane suspended
above the horizon.   Susskind, Thorlacius and Uglum\stretch proposed to take
the membrane or ``stretched horizon'' to lie only a few
fundamental lengths from the horizon (i.e the stretched horizon is very
anlogous to 't Hooft's brick wall), and postulated that the Hawking
Bekenstein entropy can be understood in terms of degrees of freedom
living on this membrane.  They further suggested that the process of Hawking
evaporation
could be completely understood in terms of interactions between the
external world and these degrees of freedom, and that the resulting S
matrix would be unitary.

't Hooft's calculation shows that this suggestion makes no sense in
quantum field theory.  The number of Schwarzchild degrees of freedom per
unit area of the membrane will be infinite in field theory, and will be
much larger than $e^{M^2 \over M_P^2}$ for any cutoff that keeps all the
degrees of freedom necessary to describe Hawking radiation.
We have discussed previously an even more serious (since it does not
make reference to very high energies or special coordinate systems)
field theoretical
argument that information cannot get out of a black hole with the Hawking
radiation.
This argument invoked the tensor product structure of field theoretic
Hilbert spaces, causality, and the smoothness of the spacetime metric on
a certain spacelike hypersurface.  In field theory, if we tried to associate
degrees of freedom with a stretched horizon, the Hilbert space would (by
locality)
be a tensor product of an inside space an outside space and a horizon space.
We could make conclusions about information carried by the infalling particles
by examining the inside space alone, and use the tensor product structure to
make negative conclusions about the possibility of that information being
carried by the horizon.

I would like to present two more
arguments (due to Susskind) that the membrane picture cannot work in quantum
field
theory.  To do so we consider a black hole of extremely large mass, or what is
the same thing, Rindler space.  The spacetime curvature is completely
negligible for
our considerations.  We know that Rindler space is simply a wedge of ordinary
Minkowski space.  We want to consider observations made by a Rindler observer
following
a timelike trajectory whose proper distance from the Rindler horizon is a few
multiples of
the fundamental length.

Now let us consider throwing a wave packet of particles at the horizon.  The
wave packet
is assumed to be localized in the transverse direction at infinity.  How will
such a packet
appear to our Rindler observer hovering just above the horizon, according to
the rules of
quantum field theory?  To some extent this is a difficult question because it
involves
scattering cross sections at very high energies and low momentum transfers.
This
regime is not well understood even in weakly coupled field theories.
We are interested
in the question of whether the cross section grows with energy, and in how
much information
is contained in the growing cross section.  A rigorous answer to the question
of
growth of the cross section is not known, but what we do know fairly
rigorously, both from
field theoretic studies in models with spin one gauge bosons,
and hadron phenomenology\foot{Here we are extrapolating
the behavior of hadrons at energies $\sim 1$$TeV$ to energies many orders of
magnitude above the
Planck scale $\!$} is that the growing part of the cross section is more or
less
universal.  In parton model language, the quantum numbers that distinguish
hadrons from
one another are carried by the valence partons, which carry finite fractions
of the longitudinal momentum of the hadron in the infinite momentum frame
\foot{A rather appropriate
frame for studying super-Planck energy collisions near the horizon $\!$}.
These behave like normal
particles with transverse wave functions which do not grow with energy, but
remain localized in the
transverse plane.  Their wave functions also Lorentz contract in a normal
manner and are thin pancakes
in the infinite momentum frame. The bulk of the high energy cross section is
due to {\it wee partons}, which have
a ${dx\over x}$ distribution in longitudinal momentum and logarithmically
spreading wave functions
in the transverse plane.  In hadron physics, the wee parton distribution is
universal and contains
no information about which hadrons are scattering (except perhaps whether
they are baryons or mesons).
We emphasize that the existence of growing cross sections and wee partons is
a conjecture in field theory.
If they do not exist then our argument is even simpler.  The important point
is that they do not carry
(much) information even if they exist.

The implications of this behavior for our Rindler observer are clear
\foot{Actually, the statements
made above apply to an inertial observer who happens to have the same
velocity as the Rindler
observer at the moment he makes his measurements.  If the measurements are
sufficiently localized
in space and in infinite momentum frame time there is probably not very much
difference
between these two observers in quantum field theory.  The distinction might
be more important
in string theory, where we do not really know how to solve the theory in
a noninertial reference frame. One extremely important aspect of the
situation that cannot be understood in terms of inertial frames is the
difference between the experiences of the inertial and accelerated observer.}.
He will find that the the information
in an infalling particle wave packet remains localized in the transverse plane.
Furthermore if he restricts his attention to the region outside a stretched
horizon which is some fixed
distance from the horizon, he will soon find that he can not measure anything
about the
infalling state at all.  The flat pancake which carries all the information
falls below the
stretched horizon in a time short compared to the Hawking evaporation time of
the black hole.
Thus, according to the rules of quantum field theory,
it is impossible for a Rindler observer to imagine that the information
carried by infalling particles gets smoothly spread over the stretched horizon
, to be emitted
later as isotropic Hawking radiation.

Susskind points out that in string theory the situation is quite different.
Growing cross sections
are built into string theory, since it is Regge behaved and the intercept of
the graviton trajectory
is $2$.  A very picturesque way of understanding this has been developed by
Karliner Klebanov and Susskind\ref\klebsuss{M.Karliner, I.Klebanov,
L.Susskind, {\it Int. J. Mod. Phys.}{\bf A3},(1988),1981.}
who did Monte Carlo simulations
of the distribution of strings predicted by the wave functions of free
bosonic string theory in light cone gauge.
 Consider a small box of fixed size in the transverse plane, and a string
whose center of mass is in that
box.  KKS ask how much of the actual length of the string is in that box.
The answer depends on a cutoff that they
imposed on longitudinal momentum.  Remembering that low longitudinal
momentum means large light cone energy
, we see that this is a cutoff on the time resolution of the observer looking
at the string.
Remember also that in light cone gauge string theory, a cutoff on
longitudinal momentum is a spatial world sheet
cutoff, a cutoff on the number of modes of the string.  KKS find that as the
cutoff is taken to infinity,
the proportion of the string that is in the box goes to zero.  This is a
symptom of the logarithmic spread of
the string in the transverse plane.  The region (measured in string
units) in which the string is confined grows logarithmically with
the cutoff.

To apply this to the black hole problem, note that for a Schwarzchild
observer supported near the horizon, watching a string fall into
a black hole a time $T$ (long) after the black hole is formed, the Lorentz
boost
between the observer's frame and the infalling string frame corresponds
to a time dilation\foot{Here we are
treating the Schwarzchild frame as a highly boosted inertial frame, because
we do not know how to do
unitary string quantum mechanics in noninertial frames.  This is a weak point
in Susskind's arguments that deserves
investigation.}$e^{T M_P^2\over 4M}$ .
Thus in order to see what is going on in the string frame the observer needs
an exponentially fine time resolution,
corresponding to a spread of the string over an area $ {T M_P^2 \over
4M M_S^{2} }$.  Thus,
in the Hawking evaporation time $T\sim {M^3\over M_P^4}$
the string has spread over an area ${M^2 \over M_P^2 M_S^2}   $.  There is an
important factor of the string coupling in these formulae
corresponding to the fact that the natural length scale for string
fluctuations is larger that the Planck length
by a factor of $1\over g$.  This ensures that the string has spread over an
area larger than the horizon, or rather that
it has been able to cover the horizon many times.  In higher dimensions the
string spreading, evaporation time, and horizon
area scale differently and the string has spread over regions bigger
than the horizon by powers of the black hole mass\foot{In all of these
considerations, it is important to take the string coupling to be
extremely weak.  The region in which our description of string
scattering is reliable is ${g^2 \over M_S^2} E < b$, the impact parameter.
In order for the energy to be high enough for string spreading to cover
the horizon, we need $ln ({E\over M_S}) = {g^4 M^2 \over M_S^2}$.
However we also must be discussing experiments with impact parameters
much less than the Schwarzchild radius of the black hole.  Thus, we must
have ${M\over M_S} \gg 1$ and $g \ll \sqrt{M_S \over M}$.  This
inequality is not satisfied by the string coupling in the real world.
Thus Susskind's estimates are really valid only in an imaginary world
with very weak coupling. I thank E. Martinec for a discussion of this
point.}

Another important difference between string theory and field theory is that
the information which distinguishes between states
is carried not by {\it valence partons}, which are localized excitations, but
by vertex operators which are conformal fields
smoothly spread over the fluctuating string.  (Examples are the Kac-Moody
currents which carry gauge quantum numbers.)
The well known inability of string theory to reproduce localized form factors
for hadrons is a symptom of this effect.
This implies that the localized Schwarzchild observer will not be able to
conclude that the information
carried by an infalling string is localized near its center of mass.  A
collection of observers spread over the horizon
would be necessary to extract the information (and they would then probably
completely change the state of the string).
Thus there is no contradiction with the claim that the information can be
emitted as isotropic thermal radiation.

In a similar vein, the Regge behavior of string scattering amplitudes and
their growing cross sections, imply that
strings do not Lorentz contract to sizes smaller than the string length.
Thus, if we put a stretched horizon
within a string length of the horizon, strings will never appear to
``fall through it'' from the point of view of a Schwarzchild
observer.  Thus in string theory, as opposed to quantum field theory, the
picture of information carried by
a stretched horizon which interacts with the outside world is not ruled out.

In assessing Susskind's claims about string theory and black holes, it is
important to understand that he is not proposing to change the picture of
physics as seen by an infalling observer near a large mass black hole.  The
spacetime geometry is smooth and so is the observer's coordinate system, and
no string theoretic
corrections to a semiclassical field theory description are important.
This is not the case for the Schwarzchild observer.  We have recalled that
in classical general relativity the Schwarzchild observer can access all of
the information
about objects that long ago fell into the black hole, but only by making
observations
on extremely short distance and time scales.  She can attribute all
interactions of
the black hole with the exterior to a membrane on the stretched horizon, but
unless she takes
her stretched horizon extremely close to the true horizon and measures time
with extreme precision, she can only discuss average properties of the hole.
It is to describe these extremely precise measurements of short times and
distance scales
that string theory, or some other theory of the small scale structure of the
world, becomes necessary.
Susskind's string theory of black hole thermodynamics is, like the classical
membrane paradigm,
a description of the physics that is tied to a particular coordinate frame
(and those related to it by
smooth coordinate transformations).

We now come to what most researchers consider the most important argument
against a unitary S-matrix,
the analysis of the spacelike surface we called $99$ in section 2.
The refutation of this
argument goes right to the heart of the difference between quantum field
theory and string theory.
The essence of the argument is that the state of the system on this spacelike
hypersurface
lives in a Hilbert space which is a tensor product of an ``inside'' space and
an ``outside'' space.
If we wish, we can add a ``membrane'' space intermediate between the two, but
this changes nothing.
We claim that we know from the prior history of the system that the inside
state is correlated with
the outside.  Then, from knowledge that the inside state has not changed very
much
 (``nothing much happens to the infalling observer
as he falls through the horizon of a large black hole'') we conclude that the
outside
density matrix has a large entropy.  When we try to rerun this argument in
string theory,
we run into an immediate snag.  Consider ordinary light cone gauge string
field theory
and let us try to divide space into two along one of the transverse
directions.  We want
to consider the entire Hilbert space as a product of an $x_1 > 0$ and and
$x_1 < 0$ Hilbert space.
We cannot.  The full Hilbert space contains states which are created from the
vacuum by
creation operators for strings which straddle the boundary.  These do not
belong in either
the plus or minus Hilbert space.  The tensor product structure of the Hilbert
space of any quantum field
theory is completely unrecognizable here.  For a black hole spacetime,
these non field theoretic states will be strings ``stranded on the
stretched horizon'',

String field theory in a half space is in fact a complicated interacting
theory. In addition to states of strings completely within the half
space, it contains the stranded states described above.  And even when
the string coupling vanishes, the half space theory contains complicated
interactions in which a closed string is annihilated and replaced by an
arbitrary number of stranded strings.  These are required to reproduce
the dynamics of free strings moving in the full space, which cross
the boundary (Fig. 9).

\ifig\fnine{Stranded Strings Violate the Locality Postulate of Field Theory,
But Can Only Be Seen With Very Short Time Resolution}
{\epsfysize=6cm \epsfbox{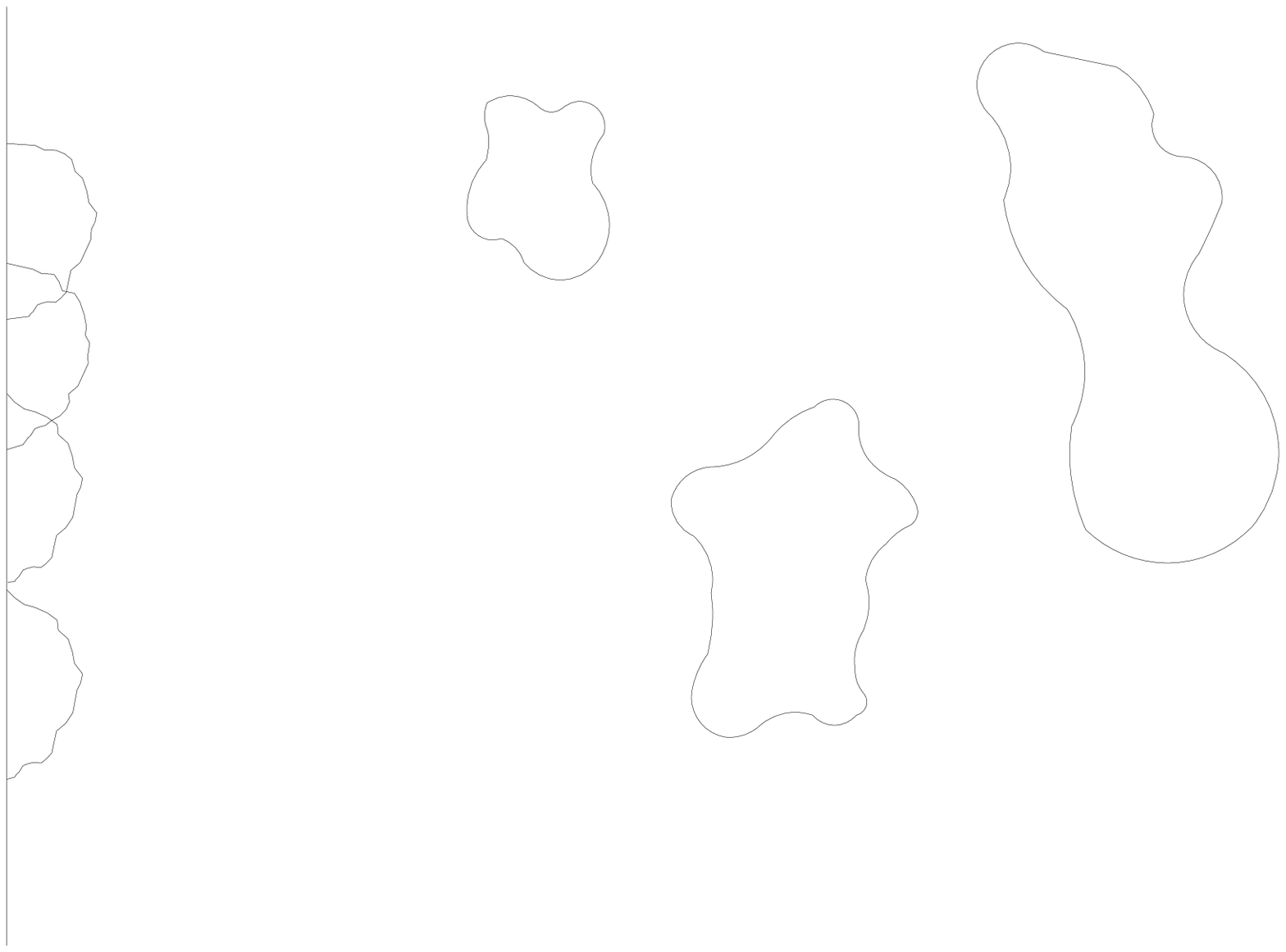}}

It is important to note that the wave functionals of even low energy states
of the string have nonzero
amplitudes for finding large strings in them.  Thus the creation operator for
a photon state with
center of mass coordinate arbitrarily far from
 the boundary has a nonzero piece which
creates stranded strings.  It is only when we agree not to study the photon
state on short time
scales that we are able to ignore this.  The large strings which appear with
finite amplitude in the photon
wavefunctional all have large world sheet wave numbers.  When we consider
averages over
light cone time intervals much larger than the string scale we can construct
an effective theory
in which string degrees of freedom with large world sheet wave numbers are
integrated out.  The only large
strings which appear in this effective theory are large smooth strings.
Neither the photon wavefunctional nor
that of any low energy excitation of the string has any significant amplitude
for such large smooth strings.
Thus, the stranded string states in the full string field theory Hilbert
space can be ignored
if we are making low energy observations of low energy states.

The situation is quite different for
observations with very short time resolution.   Consider a wave packet of
photons
 travelling in the negative $x_1$ direction, the center of which is at some
large negative $x_1$.
The string wavefunctional for this photon has finite amplitudes for string
configurations which
wander arbitrarily far from the center of the wave packet , even into the
region where $x_1 > 0$.
These are configurations which contain large contributions from very high
world sheet wave numbers, and
they oscillate with very high frequency.  However, measurements with
sufficiently fine time resolution
in the $x_1 > 0$ region should be able to discern them.  In principle
these measurements could determine the state of a photon whose center
of mass was arbitrarily far away.

Susskind thus argues that much of the information (indeed he claims all
the information) that in field theory is lost behind the horizon of the
black hole, is encoded in string theory in the state of the stranded
strings on the stretched horizon.
It would be a remarkable consistency check of these ideas if one could
calculate the entropy of stranded strings in a black hole background and
find that it agreed with the Bekenstein Hawking entropy.
Unfortunately,
things are not so simple.
Firstly, we do not know how to formulate a Hamiltonian version of string
theory in a black hole background, nor even in noninertial coordinates
in Minkowski space.  Secondly, even the problem of the
entropy of a half space in inertial coordinates\foot{This problem is of
interest because in quantum field theory formal arguments indicate that
the half space entropy is exactly equal to the Rindler
entropy\srednicketal .} is, in string theory, quite complicated.  As
noted above, it is an interacting string field theory of closed and open
strings even when the string coupling is set to zero.
Finally, there is a puzzle about the detailed form of the BH entropy
formula in any theory in which the fundamental length is larger than the
Planck length by a factor of the dimensionless coupling that controls
semiclassical expansions.

In particular, in string theory, the fundamental unit of
length is the string length $l_S \sim\sqrt{\alpha^{\prime}}$, related to
the Planck length by the dimensionless string coupling, $g_S l_S \sim
l_P$.  Thus, in terms of the natural units of string theory, the
Bekenstein Hawking entropy is an effect of order ${1\over g_S^2}$.  Any
straightforward calculation of the quantum entropy of strings in a black
hole background will, to leading order, be independent of $g_S$.
Indeed, this would seem to be a property of any quantum theory.  In the
semiclassical expansion entropy arises by tracing over states which are
small fluctuations about a classical background.  To leading order,
their description is independent of the coupling, and one would expect
the expansion of the entropy to begin at order $g^0$.

Susskind proposes to avoid this apparent difficulty by the following
heuristic, though rather convincing, line of argument.  Imagine doing
string theory in a light cone frame moving parallel to the horizon.
In this gauge we can, as above, try to make sense of local quantities in
string theory.  In particular, Susskind proposes to take over from
local field theory the notion that an observer in this frame localized
a distance $d$ from the horizon, effectively sees a bath of strings at a
$d$ dependent temperature, approaching the Hagedorn temperature as $d
\downarrow l_s$.  If this is the case, then the string partition
function at genus one will blow up in a black hole background.  It has
long been believed that in the finite temperature case,
this divergence is the signal of a phase
transition to a new phase of string theory.  In particular,
Atick
and Witten\ref\Atick{J.Atick, E.Witten, {\it Nucl. Phys.}{\bf B310},(1988),
291.} have argued that this phase transition is of
first order.  Formally, the precursor of the transition is a tachyonic
mass for certain winding
mode of strings around the compact time dimension of a finite
temperature cylinder.  Atick and Witten argue that string interactions
stabilize this instability, leading to a finite condensate of winding
modes which contributes to the free energy at order ${1\over g_S^2}$.
Susskind proposes to identify the black hole analog of the genus zero
entropy of Atick and Witten with the Bekenstein Hawking entropy.

It is beyond the reach of present stringy technology to calculate the
Atick-Witten entropy at genus zero, much less Susskind's profound
generalization of it.  Nonetheless, taken as a whole, Susskind's
proposal is, in this author's opinion, the first serious attempt to
explain the BH entropy in terms of a quantum mechanical sum over states.
His whole picture, if viewed as the quantum mechanical analog of the
Membrane Paradigm, is consistent with everything we know about black
holes and strings.  In particular, it does not contradict the use of
effective field theory for the description of the vicinity of the
horizon of a large
black hole as viewed by an infalling observer.  Classical general
relativity tells us that a Schwarzchild observer with fine enough
resolving power is able to retrieve all information about things that
fell into the black hole in ``the remote past''\foot{as seen by an
infalling observer.  Of course all these measurements are made on a
single Schwarzchild time slice.}.  It is only for the quantum mechanical
analysis of these super-Planckian measurements that one needs to resort
to string theory.

Viewed in this manner, the BH entropy formula hints at a very
interesting conclusion.  In classical physics, we can throw an infinite
amount of information into a black hole, and the Schwarzchild observer
can measure it.  The BH formula suggests that this infinity is cutoff
by quantum effects.  Bekenstein has tried to argue
\ref\bekbound{J.Bekenstein, {\it Phys. Rev.}{\bf D7},(1973),2333,
{\it Phys. Rev.}{\bf D9},(1974),3292,
{\it Phys. Rev.}{\bf D49},\break (1994),1912.} that
this cutoff reflects a fundamental quantum gravitational restriction on
the amount of information that can be contained inside a volume bounded
by an area $A$.  As far as I know, all such arguments implicitly assume
that the volume enclosed is finite, and certainly do not take into
account geometries with horizons which recede off to infinity.  The
cornucopion scenario is in direct contradiction with Bekenstein's
bounds, if they are taken to refer to the entire entropy of entanglement
of the external world with the world behind the black hole horizon.
If instead, they are taken to represent only the entropy of states near the
horizon that are in causal contact with the external world, the contradiction
is removed.  I believe that this is all that is necessary to prevent
violations of the second law of thermodynamics in the presence of black holes.

Instead, in view of Susskind's picture of the origin of the BH formula,
I would ascribe the finiteness of BH entropy to a limitation on the
information carrying capacity of stranded strings on the stretched
horizon.  Consider a string state whose center of mass is thrown into
the black hole.  In free string theory it leaves behind stranded strings
on the horizon.  Once interactions are taken into account, these
stranded strings
can break, combine with other strings, and in this way, ``lose contact''
with the original state which deposited them.  Information about the
infalling state is now truly lost to the stranded strings and thus to
the external observer.  The horizon is thus a semipermeable membrane for
information:  that information represented by the BH entropy is
information about the state of the stranded strings on the stretched
horizon., but it is not complete information about what fell into the
black hole.  The idea that complete information is in principle
accessible is an artifact of classical physics, corresponding to the
divergence of the BH entropy in the $g_S \rightarrow 0$ limit.

This lecture has been only a brief summary of Susskind's ideas.  I have
stressed
primarily those points where I felt that a clearer explanation than
could be found in the literature was necessary.  For more details, I refer
the reader to Susskind's original papers.
\newsec{Conclusions}
I believe that by combining the notion of cornucopions with Susskind's ideas
about string theory and BH entropy, we have for the first time the outline
of a sensible story about the problem of information loss in black hole
evaporation.
These two seemingly contradictory aspects of the description are the analog
of the infalling and supported observers' points of view
in classical general relativity.  The cornucopion scenario tells us what
is "really going on" behind the horizon.  It cannot easily account for
the BH entropy, which is a notion relevant to the observations of the
external observer\foot{Note that in the interpretation advanced in these
lectures,
the BH entropy cannot be identified with the entropy of entanglement
of the interior and exterior of the black hole.  The latter is an S-matrix
concept, which can be defined in an observer independent manner.  If our
interpretation is correct, the thermodynamic BH entropy is a concept that
is useful for the external observer only. }.  Susskind's interpretation of
the BH entropy is a concrete realization of the heuristic treatment of
this entropy by the membrane paradigm.  It is remarkable that we have to
give up the rules of quantum field theory in order to find a consistent
realization of these semiclassical ideas.  It is even more remarkable
that string theory seems to provide the required generalization of field
theory.  If these arguments are verified, 't Hooft's bold claim that the
resolution of the paradoxes of Hawking radiation would lead us to the
correct theory of quantum gravity will be vindicated.

Taken by itself, Susskind's picture could be advanced (and Susskind so advances
it)
as proof that {\it all} of the information in the black hole is returned
to the external observer.  As indicated above, I do not think that this
is necessarily correct.  All of the recent work on remnants has led to a
picture
in which black hole formation and
evaporation terminates, at least for black holes of sufficiently
large magnetic charge, in a semiclassical spacetime in which the number of
different causally disconnected asymptotic regions is different in the future
than in the past.  Basically, this is the picture of a black hole
that comes out of classical general relativity, with only the singular behavior
of that theory removed.  I believe that perturbative string theory on
such a spacetime would lead to the prediction that the S-matrix for the
original asymptotic region is non-unitary (basically, because even though
string theory is nonlocal, it satisfies the cluster property of the S-matrix).
\foot{Of course, the spacetimes in question are not solutions of the
classical equations
of string theory, so naive perturbative string theory is not quite applicable.}

It is somewhat of a jump, (and I think at odds with a lot of semiclassical
reasoning about what happens to freely falling observers in regions of
small spacetime curvature) , to conclude from the fact that string theory may
explain the Bekenstein-Hawking entropy in terms of degrees of freedom
connected with the horizon, that {\it no} information is lost in black hole
evaporation.  In the present state of our knowledge (and even accepting
Susskind's claims about stranded strings as fact) one may equally well
argue that the BH entropy formula simply represents a quantum string theoretic
limit on the information carrying capacity of the degrees of freedom
connected to the horizon.  Perhaps this limit arises because (due to string
interactions)
the stranded strings that were originally connected to states whose center of
mass fell through the horizon, can break, and lose contact with the
information they originally carried.  In this interpretation,
the BH formula represents only that portion
of the entropy of black hole formation
that remains causally connected to the external observer.
Such an interpretation seems consistent with the semiclassical
description of the observations of an infalling observer who sees the
formation of a new asymptotic region of spacetime.

On the other hand, someone
who insists on a unitary S-matrix for the world outside a black hole must
either deny the possibility of the formation of causally disconnected
asymptotic regions of space or insist on a radical reformulation of our notions
of space-time and/or quantum mechanics.  Such a point of view is certainly more
exciting than the conservative approach that I have advocated.  It leads
to the notion of ``holographic'' theories of spacetime in which the number
of states of a space with $D$ dimensions is a finite multiple of the area
of a transverse spatial slice like a black hole horizon
\ref\holograph{G. 't Hooft, gr-qc 9310006; L. Susskind, hep-th 9409089 }.
There are even indications that string theory is such a holographic theory
\ref\stringholo{L. Susskind, op. cit.; C. Thorn,
{\it Reformulating String Theory With the ${1\over N}$
 Expansion}, \break Talk Given at the First International A.D.Sakharov
Conference, Moscow,
Russian Federation, May 1991.; S.Das and A. Jevicki, {\it Mod. Phys. Lett.}
{\bf A5},(1990),1639.}.  At the moment however, the intermediate position
that I have outlined
seems more plausible to me.

And this brings us back to the frustration with which I began this talk.
Black hole evaporation is a purely theoretical subject.  The only aspect
of my discussion that is possibly accessible to experimental test (and that
possibility is quite remote), is the claim that there exist pointlike
cornucopions that behave like macroscopic classical objects.  Apart from
that, the resolution of the controversy surrounding Hawking radiation
appears to await the construction and qualitative solution of a mathematically
consistent theory of quantum gravity.  It is still possible for an honest
researcher to uphold any of the three major points of view about the
outcome of the controversy (though personally I feel that theories of true
information loss are probably not consistent).

The avenue for future progress appears to branch into three rocky paths.
The first is to pursue the study of remnants in string theory, and to try
to derive from string theory the nonsingular models of remnants that have
been constructed by hand.  In particular, the question of whether there is
a threshold for excitation of the continuum modes of a cornucopion
should be pursued vigorously, as well as the question of the number of
cornucopions expected in our horizon volume.  These questions bear on the only
experimentally accessible aspect of our subject.  The second path is the
verification of the 't Hooft-Susskind picture of the states that are counted
by the Bekenstein-Hawking entropy of a black hole\ref\atish{A.Dabolkhar,
hep-th 9409158.}.  I deliberately split off from
this a third path: the study of holographic theories, that is theories
in which the S-matrix of the external observer is exactly unitary.
In particular one
wants to ask of such theories how the unitarity of the external S-matrix
of a black hole can be made consistent with the apparent feeling of an
infalling observer that there are {\it asymptotic} states causally
disconnected from
the outside.  Although string theory appears to exhibit nonlocality, it does
seem to satisfy the usual rules of cluster decomposition of the S-matrix, and
to preserve the solutions of the Einstein equations for large smooth
geometries.
For large mass black holes, there seems to be a clear indication of the
existence of new asymptotic states, causally disconnected from the world
outside the horizon.  Indeed, as Susskind often emphasizes, we have no
evidence that
we are not living inside of a black hole\foot{In most inflationary models of
the universe, this is certainly the case.}.
If string theory is holographic, and the external S-matrix unitary, it is
very hard to understand how to interpret the experience of ``internal''
observers.

At present, all of these paths seem steep and rocky and difficult to climb.
Those who follow them will surely want to console themselves with dreams of
the fantastic vistas
that will open up before them when they reach the top.
\vfill\eject
\newsec{\bf Acknowledgements}
Conversations about black hole physics with G. 't Hooft, E. and H. Verlinde,
S.Das, and E.Martinec are gratefully acknowledged.
Much of the content of these lectures was written while the author was
in residence at the Institute for
 Theoretical Physics in Santa Barbara during the spring semesters
 of 1993 and 1994.  I would like to thank J.Langer, M.Stone and
the Institute staff for providing me with a wonderful place to get some work
done.  Conversations with M.Srednicki, S.Giddings, and expecially A.Strominger
were invaluable to me, and without them these lectures would be much less clear
than they are.  I would also like to take this opportunity to thank my
collaborators, A.Dabolkhar, M.Douglas, and especially M.O'Loughlin for hours of
enjoyable debate about the subject of black holes and information loss.
My colleague S.Shenker deserves his usual acknowledgement for penetrating
insights that have opened up whole new areas of investigation for me.
The lectures were completed at the Aspen Center for Physics in the summer
of 1994 and the Laboratoire de Physique Theorique of the Ecole Normale
Superieure in Paris, in the fall.  I would like to thank all of the members
and staff
of these institutions for their hospitality.  I.Klebanov and R. Dijkgraaf
also deserve a vote of thanks for organizing an excellent Spring School
and twisting my arm to attend it.  Finally, I would like to thank L.Susskind
for many conversations about physics, and for continually challenging us all
with his original insights.  This work was supported in part by the
Department of Energy under grant number DE-FG05-90ER40559.
\vfill\eject

\listrefs

\end